\documentstyle[prl,aps,draft,epsf,twocolumn]{revtex}

\def\be{\begin{equation}}
\def\ee{\end{equation}}
\def\bea{\begin{eqnarray}}
\def\eea{\end{eqnarray}}
\def\nn{\nonumber\\}
\def\r#1{(\ref{#1})}

\def\({\left(}
\def\){\right)}
\def\X#1{\Omega_{#1}}
\def\HDM{{\cal H}_{\text{\tiny DM}}}

\begin{document}

\title{Finite-temperature dynamical magnetic susceptibility of
quasi-one-dimensional frustrated spin-1/2 Heisenberg antiferromagnets} 

\author{Marc Bocquet$^1$, Fabian H.L. Essler$^1$, Alexei
M. Tsvelik$^2$ and Alexander O. Gogolin$^3$}
\address{$^{1}$ Department of Physics, Warwick University, 
Coventry CV4 7AL, UK}
\address{$^{2}$ Department of Physics, 
        University of Oxford, 1 Keble Road, Oxford OX1 3NP, UK}
\address{$^3$ Department of Mathematics, Imperial College, 
180 Queen's gate, London SW7 2BZ, UK}
\maketitle
\begin{abstract}
We study the dynamical response of frustrated, quasi one dimensional
spin-1/2 Heisenberg antiferromagnets at finite temperatures. We allow
for the presence of a Dzyaloshinskii-Moriya interaction.
We concentrate on a model of weakly coupled planes of anisotropic
triangular lattices.  
Combining exact results for the dynamical response of one
dimensional Heisenberg chains with a Random Phase Approximation (RPA)
in the frustrated interchain couplings, we calculate the dynamical
susceptibility in the disordered phase. We investigate the instability
of the disordered phase to the formation of collective modes. We find a
very weak instability to the formation of incommensurate magnetic
order and determine the ordering temperature and wave vector. We also
determine the effects of uniform magnetic fields on the ordering
transition. 
\end{abstract}

\maketitle

\section{Introduction}

A defining property of quasi one dimensional (1D) magnets is the weakness
of the interchain coupling, whose presence nevertheless usually leads
to three-dimensional ordering. It is natural to consider the ratio of the
transition temperature $T_c$ to the bandwidth of the excitation spectrum
$D$ as a quantitative measure of one dimensionality. In systems where 
$T_c/D \ll 1$ the ordering transition occurs in a state where spins on
each chain are already highly ``collectivized''. This means that a
very significant proportion of the spectral weight is concentrated in a
piece of the spin-spin correlation function, which is essentially
one dimensional \cite{Tennant95a,Tennant95b,bella,Coldea00}. 
This suggests that the most natural approach to the problem of phase
transitions in such systems is to treat them as instabilities
\cite{BY}, driven by weak interchain interactions, of an ensemble of
1D chains. This approach utilizes the knowledge of the correlation
functions of individual 1D chains, obtained by various non-perturbative
methods. In this way one will automatically reproduce a distinct
feature of quasi-1D magnets, namely the presence of a broad incoherent
continuum in the dynamical structure factor. Hence this approach is
quite different from the conventional spin wave theory, which uses
individual spins as elementary building blocks. Spin wave theory is
known to work well for the coherent (single-particle) parts of the
spectra, but it is notoriously difficult to obtain the incoherent
parts within this approach. Therefore it works poorly in one
dimension, especially for spin-1/2 magnets, where the incoherence is
very strong. The latter  is due to the fact that excitations of a
spin-1/2 Heisenberg chain carry quantum numbers different from those
of spin waves.

Some of us have already applied the approach based on weakly coupled
chains to describe the ordered phase of cubic lattice quasi-1D
antiferromagnets \cite{Delfino} (see also \cite{Sandvik99,schulz}).
In the present work we apply an
analogous method to the disordered phase and the ordering transition
in quasi-1D, frustrated  spin-1/2 antiferromagnets with
Dzyaloshinskii-Moriya (DM) interaction. The effects of frustration on
quantum magnets are very interesting and continue to attract much
experimental
\cite{Coldea00,Coldea96,Coldea97,Coldea97b,oleg,srcuo2} and
theoretical \cite{frustration,rrp,Weihong99,Chung00,rod,NGE,zz}
attention.

We focus on calculating the dynamical magnetic
susceptibility $\chi(\omega,\vec{k})$, which is related to the
dynamical structure factor $S(\omega,\vec{k})$ relevant for inelastic
neutron scattering experiments by 

\begin{equation}
S^{\alpha\beta}(\omega,\vec{k})=-\frac{1}{1-\exp(-\omega/T)}\ {\rm
Im}\chi^{\alpha\beta}(\omega,\vec{k}). 
\label{DSF}
\end{equation}
Note that throughout this paper we will mainly present plots of $-{\rm
Im}\chi(\omega,\vec{k})$ rather than of the structure factor itself; the
trivial prefactor in (\ref{DSF}) can be easily restored if desired.

\section{The model} 
\begin{figure}[h]
\begin{center}
\epsfxsize=5cm
\epsfbox{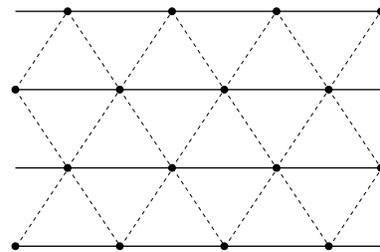}
\caption{Exchange paths within the planes: solid lines denote the
strong exchange $J_\parallel$, dotted lines the weaker, frustrated 
exchange $J_\perp$
.}
\label{fig:lattice}
\end{center}
\end{figure}

Our work is inspired by the recent experiments on the frustrated
spin-1/2 magnet Cs$_2$CuCl$_4$\cite{Coldea00,Coldea96,Coldea97,Coldea97b}. 
It was suggested in \cite{Coldea00,Coldea97b} that Cs$_2$CuCl$_4$ is
described by a Heisenberg model on weakly coupled planes of
anisotropic triangular lattices. Within the planes the exchange
couplings are as indicated in Fig.\ref{fig:lattice}.

Neighbouring planes are coupled by a weak antiferromagnetic exchange
$J^z$. In Cs$_2$CuCl$_4$ neighbouring planes are slightly shifted with
respect to one another, but in order to keep things simple we ignore
this shift. The effective magnetic Hamitonian is thus given by
\bea
{\cal H}&=&\sum_k{\cal H}_{\rm plane}^{(k)}+{\cal H}^{(k,k+1)}_{\rm interplane}\ ,\nn
{\cal H}_{\rm plane}^{(k)} &=&J_\parallel\sum_{i,j}
\vec{S}_{i,j,k}\cdot\vec{S}_{i+1,j,k}\nn
&&+J_\perp\sum_{i,j}\vec{S}_{i,j,k}\cdot\left[\vec{S}_{i,j+1,k}
+\vec{S}_{i-1,j+1,k}\right] ,\nn
{\cal H}^{(k,k+1)}_{\rm interplane}
&=& J_z \sum_{i,j}\vec{S}_{i,j,k}.\vec{S}_{i,j,k+1} \, . 
\label{H}
\eea
In the present work we analyze the model (\ref{H}) in a quasi-1D framework.
We choose to view the Hamiltonian (\ref{H}) as spin-1/2 chains with
exchange $J_\parallel$, coupled by a weaker, frustrating,
nearest-neighbour exchange $J_\perp$ within a plane. 
Finally there is a very weak antiferromagnetic coupling $J_z$ between
neighbouring planes. 

Experimental estimates for the exchange couplings in ${\rm Cs_2CuCl_4}$
are $J_{\parallel}=0.37$ meV, $J_\perp/J_{\parallel} \simeq 0.33 (1)$,
and $J_z/J_{\parallel}\approx 0.05$ \cite{Coldea00,radu}. Although the
interchain coupling $J_\perp/J_\parallel$ is considerable, the
smallness of the ratio of transition temperature to bandwidth $T_c/\pi
J_{\parallel} \approx 0.05$ indicates that the type of approach we are
advocating might be applicable to ${\rm Cs_2CuCl_4}$. 

In addition to the exchange interactions present in (\ref{H}) we
allow for the presence of a DM interaction. Our motivation is once
again the situation in ${\rm Cs_2CuCl_4}$, where a DM
interaction appears to be present
\cite{Coldea00,Coldea96,Coldea97,Coldea97b,radu} although it is not
straightforward to estimate its magnitude. This is because the
superexchange between two Cu spin-1/2 occurs through two Cl$^-$
ligands and not a single one. One possible DM interaction that
respects the crystal symmetry of ${\rm Cs_2CuCl_4}$ is
\be
{\cal I}_{\text{\tiny DM}} = \sum_{i,j,k} \vec{ D}\cdot\left[\vec{S}_{i,j,k}
 \times\vec{S}_{i+1,j,k}\right] .
\label{DM}
\ee
Note that this form of the DM interaction only couples spins on a same
chain.
We mainly concentrate on a DM interaction of the type (\ref{DM}) but
discuss how to treat more general forms in section \ref{sec:DMgeneral}. 

The outline of this paper is as follows. In section \ref{sec:1d} we
review results for the dynamical susceptibility of a single spin 1/2
Heisenberg chain (with exchange anisotropy). In section \ref{sec:3d}
we study the effects of a frustrated interchain coupling between
spin-1/2 chains. We determine the finite temperature dynamical
structure factor by combining the results for single chains with a
random phase approximation in the interchain couplings. We show that
an instability towards an incommensurate, ordered state develops at a
critical temperature, which we determine as well. In section
\ref{sec:H} we study the effects of a uniform magnetic field. 
In sections \ref{sec:1dDM}, \ref{sec:3dDM} and \ref{sec:HDM} we 
carry out the analogous programme for spin-1/2 chains with DM
interaction. In section \ref{sec:DMgeneral} we show how to treat
more general DM interactions within our calculational scheme. Section
\ref{sec:summ} contains a summary of our results and our conclusions.

%%%%%%%%%%%%%%%%%%%%%%%%%%%%%%%%%%%%%%%%%%%%%%%%%%%%%%%%%%%%%%%%%%%%%%
\section{Dynamical susceptibility of a single spin-1/2 XXZ chain}
%%%%%%%%%%%%%%%%%%%%%%%%%%%%%%%%%%%%%%%%%%%%%%%%%%%%%%%%%%%%%%%%%%%%%%
\label{sec:1d}

\begin{figure}[h]
\epsfxsize=8cm
\epsfbox{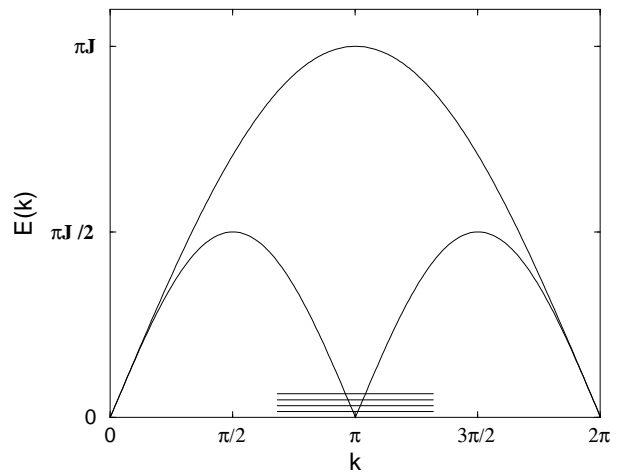}
\caption{
Two-spinon dispersion for the isotropic Heisenberg chain. Also
indicated are the sections for which we plot the dynamical susceptibility.}
\label{fig:disp}
\end{figure}

Let us first consider a single Heisenberg chain with ``$XY$-like'' exchange
anisotropy $0\leq \Delta\leq 1$
\begin{equation}
{\cal H}_{\text{\tiny XXZ}}=J_{\parallel}\sum_j S^x_jS^x_{j+1}+S^y_jS^y_{j+1}
+\Delta S^z_jS^z_{j+1}.
\label{hxxz}
\end{equation}
We will be particularly interested in the regime $\Delta\approx 1$. 
The spectrum of low-lying excitations consists of a two-spinon
continuum \cite{FT} and is shown in Fig.\ref{fig:disp} for the
isotropic case $\Delta=1$. There are gapless modes at small momentum
and at momentum $\pi$, where most of the spectral weight is
concentrated. Spinons carry quantum number $S=\pm 1/2$ and no
one-spinon excitations are present.

The large distance behaviour of the finite temperature
dynamical susceptibility can be determined by combining results
obtained from the Bethe Ansatz solution \cite{BA,vladb} of (\ref{hxxz}) with field
theory techniques \cite{boso,schulz1,Lukyanov,Affleck98,Barzykin00a}
\bea
&&\chi^{\rm xx}(\tau,x)=(-1)^x\,
A_x(\eta)\bigg[1-\(\frac{\Lambda}{T}\)^{4\eta-4}\bigg]^{1/2}\nn 
&&\times \left[\frac{\(\frac{\pi T}{u}\)^2}
{ \sinh\(\frac{\pi T}{u} (x-iu\tau)\) \;\sinh\(\frac{\pi T}{u}
(x+iu\tau)\)}\right]^{\frac{\eta}{2}} \, .
\label{chixxz}
\eea
%\left[1-\(\frac{A^2}{x^2+\tau^2}\)^{2-4\alpha}\right]
Here $\eta$ is related to the exchange anisotropy $\Delta$ by 
\be
\eta
=1-\arccos(\Delta)/\pi \label{eta}
\ee
and 
\be
u=J_{\parallel} a_0\sin(\pi\eta)/(2-2\eta) 
\ee
 is the
spin velocity. 

The nonuniversal amplitude $A_x(\eta)$ has been determined by
Lukyanov and Zamolodchikov \cite{LukZam}
\bea
&&A_x(\eta)=\frac{1}{8(1-\eta)^2}\left
[ \frac{\Gamma(\frac{\eta}{2(1-\eta)})}
{2\sqrt{\pi}\,\Gamma(\frac{1}{2(1-\eta)})}\right]^\eta \times \nn 
&& \exp\( -\int_0^\infty \! \frac{dt}{t} \Bigl[\frac{\sinh(\eta
t)}{\sinh(t)\cosh((1-\eta)t)}-\eta \, e^{-2t} \Bigr]\) \,.
\eea
Finally, $\Lambda$ is a nonuniversal scale, which recently has been
calculated in the isotropic case \cite{Barzykin00}
\begin{equation}
\Lambda/J_{\parallel}\approx 24.27\ .
\end{equation}
The third factor on the RHS of (\ref{chixxz}) stems from a renormalization-group (RG)
improvement in the leading irrelevant perturbation to the conformally
invariant scaling limit of (\ref{hxxz}) \cite{Affleck98}. There are
further logarithmic corrections to (\ref{chixxz}), some of which have
been calculated in the isotropic case \cite{Barzykin00a} and lead to
(small) corrections to some of the formulas we give below. We note that
the RG improvement in (\ref{chixxz}) was done using the inverse
temperature as the RG scale. It would be interesting to investigate
the effects of working with two scales i.e. the inverse temperature
and the Euclidean distance. The RG improvements are important only in
the vicinity of the isotropic point $\Delta=1$.

The frequency and momentum dependent dynamical susceptibility is
obtained by Fourier transformation and analytic continuation of the
time-ordered imaginary time correlation function (\ref{chixxz}) (see
\cite{schulz1,Barzykin00,Schulz98,Tsvelik95})~: 
\bea
\chi^{\rm xx}(\omega,\pi+k)&=&\Phi(T)\frac{
\Gamma\(\frac{\eta}{4}-i\frac{\omega-uk}{4\pi
T}\)}{\Gamma\(1-\frac{\eta}{4}-i\frac{\omega-uk}{4\pi
T}\)}\nn
&&\times \frac{\Gamma\(\frac{\eta}{4}-i\frac{\omega+uk}{4\pi T}\)}
{\Gamma\(1-\frac{\eta}{4}-i\frac{\omega+uk}{4\pi T}\)} \, . 
\label{chioq}
\eea
Here $\Phi(T)$ is given by
\be
-A_x(\eta) \frac{\sin(\pi\frac{\eta}{2})\Gamma^2(1-\frac{\eta}{2})}{u}
\bigg[1-\(\frac{\Lambda}{T}\)^{4\eta-4}\bigg]^{\frac{1}{2}} \bigg[\frac{u}{2
\pi T}\bigg]^{2-\eta}\, . 
\ee

The result (\ref{chioq}) is valid for momentum transfers close to the
antiferromagnetic wave number $q\approx\pi$, low energies
$\omega/J_{\parallel}\ll 1$ and low temperatures $T/J_{\parallel}\ll 1$.

The result for the isotropic point $\Delta=1$ is obtained by taking the
limit $\eta\to 1$. In this limit $A_x$ diverges and
$\sqrt{1-\(\Lambda/T\)^{4\eta-4}}$ goes to zero according to
\begin{eqnarray}
A_{x}(\eta)&\longrightarrow& \frac{1}{2(2\pi)^{3/2}\sqrt{1-\eta}}\ ,\\
\sqrt{1-\(\Lambda/T\)^{4\eta-4}}&\longrightarrow&
2\sqrt{1-\eta}\sqrt{\ln\(\Lambda/T\)} \ ,
\end{eqnarray}
so that their product tends to $\sqrt{\ln(\Lambda/T)}/(2\pi)^{3/2}$.

\begin{figure}[h]
\epsfxsize=8cm 
\epsfbox{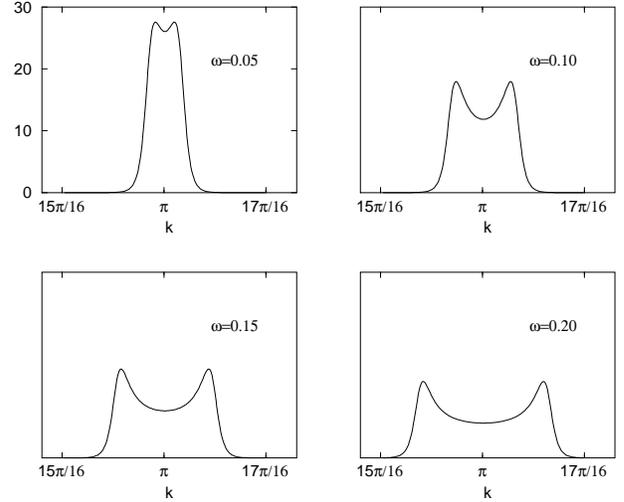} 
\caption{Imaginary part of the dynamical susceptibility $-{\rm Im}\chi^{\rm
xx}(\omega,k)$ in units of $1/J_\parallel$ for the isotropic
Heisenberg chain as a function of $q$ for four different values of
$\omega/J_{\parallel}$ at a temperature of $k_BT/J_{\parallel}=0.01$.}  
\label{fig:chi1d}
\end{figure}

In Fig.\ref{fig:chi1d} we show the imaginary part of the transverse
susceptibility in the isotropic case for the ``constant-energy scans''
indicated in Fig.\ref{fig:disp}. The two-spinon continuum is clearly
visible. The maxima at fixed frequency occur close to the boundaries
of the two-spinon continuum.

\begin{figure}[h]
\epsfxsize=8cm
\epsfbox{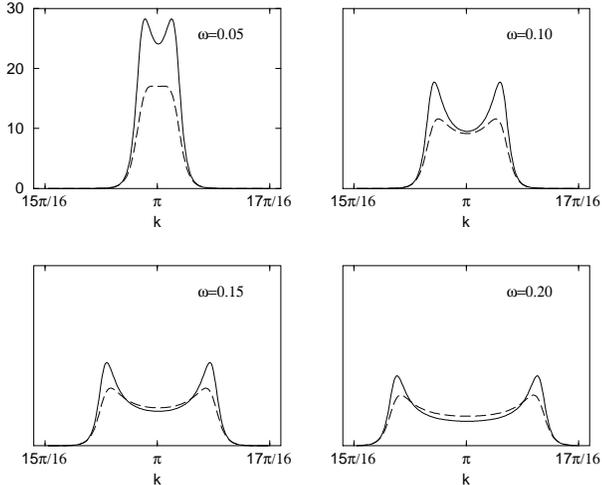}
\caption{Minus the imaginary part of the transverse (full line)
versus longitudinal dynamical susceptibility (dashed line)
in the anisotropic Heisenberg XXZ chain with
$\Delta=0.9$. }  
\label{fig:chixxz}
\end{figure}

The longitudinal dynamical susceptibility is given by
\bea
\chi^{\rm zz}(\omega,\pi+k)&=&\Phi^\prime(T)
\frac{\Gamma\(\frac{1}{4\eta}-i\frac{\omega-uk}{4\pi T}\)}
{\Gamma\(1-\frac{1}{4\eta}-i\frac{\omega-uk}{4\pi T}\)}\nn
&&\times
\frac{\Gamma\(\frac{1}{4\eta}-i\frac{\omega+uk}{4\pi T}\)}
{\Gamma\(1-\frac{1}{4\eta}-i\frac{\omega+uk}{4\pi T}\)},
\label{chi1dzz}
\eea
where $\Phi^\prime(T)$ is given by \cite{Lukyanovzz}
\bea
&&\Phi^\prime(T)=
-A_z(\eta) \frac{\sin(\pi/2\eta)\Gamma^2(1-1/2\eta)}{u}
\bigg[\frac{u}{2\pi T}\bigg]^{2-\frac{1}{\eta}}\nn
&&\quad\times
\bigg[1-\(\frac{\Lambda}{T}\)^{2\eta-2}\bigg]^{\frac{1}{2}}
\bigg[1+\(\frac{\Lambda}{T}\)^{2\eta-2}\bigg]^{-\frac{3}{2}}\ ,\nn
&&A_z(\eta)=
\frac{2}{\pi^2}\left
[ \frac{\Gamma(\frac{\eta}{2(1-\eta)})}
{2\sqrt{\pi}\,\Gamma(\frac{1}{2(1-\eta)})}\right]^{\frac{1}{\eta}} \times \nn 
&& \exp\( \int_0^\infty \! \frac{dt}{t} \(\frac{\sinh((2\eta-1)
t)}{\sinh(\eta t)\cosh((1-\eta)t)}-\frac{2\eta-1}{\eta} e^{-2t}
\)\).\nn
\eea
For $\Delta<1$ the longitudinal structure factor at low energies is
weaker than the transverse one. In Fig.\ref{fig:chixxz} we compare the
transverse and longitudinal susceptibilities for an XXZ chain with
$\Delta=0.9$. 

%%%%%%%%%%%%%%%%%%%%%%%%%%%%%%%%%%%%%%%%%%%%%%%%%%%%%%%%%%%%%%%%%%%%%%%
\section{Dynamical susceptibility for coupled Heisenberg chains}
%%%%%%%%%%%%%%%%%%%%%%%%%%%%%%%%%%%%%%%%%%%%%%%%%%%%%%%%%%%%%%%%%%%%%%%
\label{sec:3d}
Let us now take into account the interchain couplings $J_\perp$ and
$J_z$. We do this by a Random Phase Approximation (RPA) \cite{RPA,schulz},
which in the present context can be understood as the leading term of
an expansion in the inverse coordination number of the lattice. The
dynamical susceptibility of (three-dimensionally) coupled chains are of
the form 
\begin{equation}
\chi_{\text{3d}}^{\alpha\alpha}(\omega,\vec{k})= 
\frac{\chi^{\alpha\alpha}(\omega,k)}{1-2J(k,k_y,k_z)\,\chi^{\alpha\alpha}
(\omega,k)}\ ,
\label{chi3d}
\end{equation}
where $\alpha=x,y,z$ and  $\chi^{\alpha\alpha}(\omega,k)$ is the
dynamical susceptibility of a single chain. The Fourier
transform of the interchain spin-spin couplings $J(k,k_y,k_z)$ is
given by
\bea
J(k_x,k_y,k_z)&=&J_z \cos(k_z)\nn &&
+J_{\perp}\(\cos(k_y)+\cos(k_x-k_y)\) \, . 
\label{jofk}
\eea
We recall that we have assumed the coupling along the $z$-direction
(between planes) to be antiferromagnetic, i.e. unfrustrated. We can
recover the results for a weak, frustrated coupling between planes
with $J_z\ll J_\perp$ by simply setting $J_z=0$ in our
formulas. Whenever we thus refer to $J_z=0$ in the following, this should be
understood as corresponding to a weak, frustrated coupling along the
z-direction. 

In the experiments on ${\rm Cs_2CuCl_4}$ the dynamical structure
factor was measured for momentum transfers along the chain
direction. We therefore concentrate on such momentum transfers in the
framework of our approach.

\begin{figure}[h]
\epsfxsize=8cm
\epsfbox{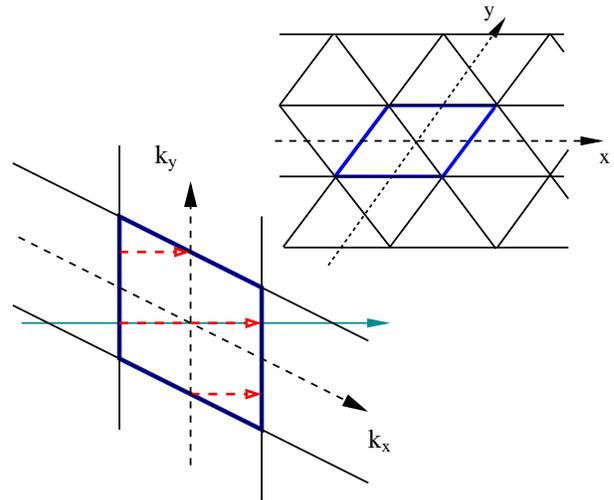}
\caption{Primitive cell of the triangular lattice and corresponding
first Brillouin zone. Wavenumbers corresponding to momentum transfer
along the chain direction are indicated by arrows.} 
\label{fig:fbz}
\end{figure}

The elementary cell and the first Brillouin zone of the planar
triangular lattice are shown in Fig.\ref{fig:fbz}. 
If one probes the magnetic properties of the crystal along the
direction of the spin chains, the wave vector transfer follows
the trajectory $k_y=k_x/2 \text{ mod } (2\pi)$ in the first Brillouin
zone and has been represented by three arrows on figure \ref{fig:fbz}. 
We see that the wave number transfer along the chain direction varies
in the interval $[0,4\pi]$. 

\begin{figure}[h]
\epsfxsize=8cm
\epsfbox{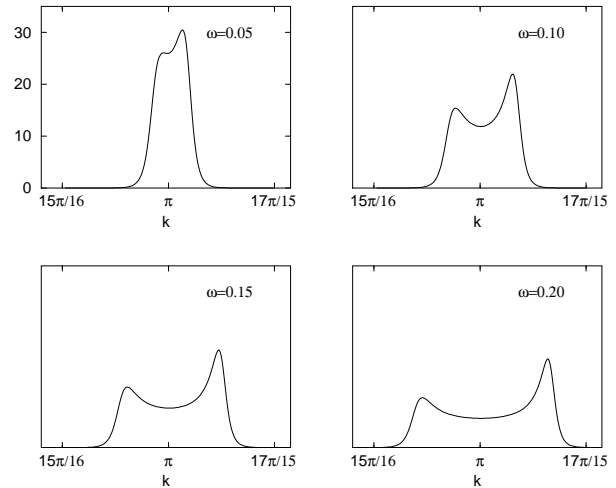}
\caption{Imaginary part of the dynamical susceptibility $-{\rm
Im}\chi^{\rm xx}_{\rm 3d}(\omega,\vec{k})$ for momentum transfers along
the chain direction. The parameters are $J_\perp/J_{\parallel}=0.1$, $T/J_{\parallel}=0.01$
and $J^z=0$. We see that the frustrated interchain coupling leads to
an asymmetric line shape.} 
\label{fig:chixx_2d}
\end{figure}

\begin{figure}[h]
\epsfxsize=8cm
\epsfbox{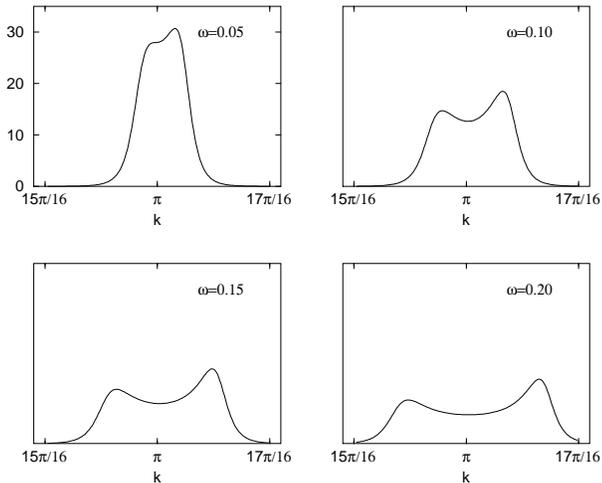}
\caption{$-{\rm Im}\chi^{\rm xx}_{\rm 3d}(\omega,\vec{k})$ for
momentum transfers along the chain direction. The parameters are 
$J_\perp/J_{\parallel}=0.1$,\hfill\break $T/J_{\parallel}=0.02$ and 
$J^z/J_{\parallel}=0.01$.}
\label{fig:chixx_3d}
\end{figure}

In Figs.\ref{fig:chixx_2d} and \ref{fig:chixx_3d} we plot (minus) the
imaginary part of $\chi^{\rm xx}_{\rm 3d}(\omega,\vec{k})$ for
momentum transfers along the chain direction for four different values
of $\omega/J_\parallel$. We have chosen parameters such that
$J_\perp/J_\parallel=0.1,\ T/J_\parallel=0.01,\ J^z=0$ and
$J_\perp/J_\parallel=0.1,\ T/J_\parallel=0.02,\ 
J^z=0.01$ respectively. In both cases the temperature is chosen to be
above the transition to an ordered state (see below) as it must for our
approach to apply.

\begin{figure}[h]
\begin{center}
\epsfxsize=7cm
\hskip .5cm\epsfbox{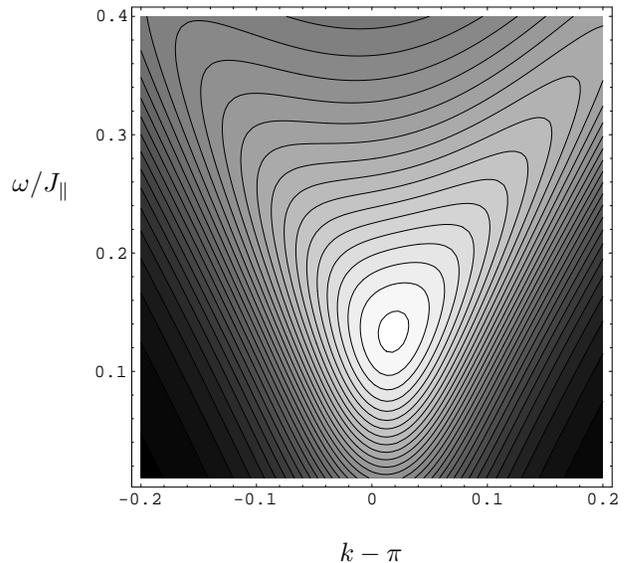}
\end{center}
\vskip-5cm ${\omega}/{J_\parallel}$
\vskip4.5cm
\hskip4cm$k-\pi$\vskip .2cm
\caption{Dynamical structure factor $S^{xx}(\omega,\vec{k})$ for
momentum transfers along the chain direction. The parameters are 
$J_\perp/J_{\parallel}=0.2$, $T/J_{\parallel}=0.05$ and 
$J^z/J_{\parallel}=0.02$.}
\label{fig:contour}
\end{figure}

Figs. \ref{fig:chixx_2d} and \ref{fig:chixx_3d} show very clearly that
the main effect of the frustrated interchain coupling is to make the
line shape {\sl asymmetric}. The susceptibility of uncoupled chains
for momentum transfer along the chain direction is symmetric around
the antiferromagnetic wave number $k=\pi$ as can be seen for example
in Fig.\ref{fig:chi1d}. This symmetry is lost when the interchain
coupling is taken into account. The asymmetry increases with
increasing $J_\perp$. The curves for $J_z=0$ and $J_z=0.01$ are
qualitatively similar. Another important feature of the imaginary part
of the dynamical susceptibility is that it looks {\sl
incommensurate} -- the maximum occurs at an incommensurate value of
the momentum transfer along the chain direction. This is most easily seen in a 
contour plot such as Fig.\ref{fig:contour}, where the dynamical
structure factor $S^{xx}(\omega,\vec{k})$ is shown as a function of
energy and momentum transfer along the chain direction.
The dynamical structure factor in the disordered phase of
Cs$_2$CuCl$_4$ displays both incommensurabilities and an asymmetric
structure \cite{Coldea00}.

%%%%%%%%%%%%%%%%%%%%%%%%%%%%%%%%%%%%%%%%%%%%%%%%%%%%%%%%%%%%%
\subsection{Instability condition and critical temperature}
%%%%%%%%%%%%%%%%%%%%%%%%%%%%%%%%%%%%%%%%%%%%%%%%%%%%%%%%%%%%%

The RPA expression for the dynamical susceptibility is valid in the
disordered phase, i.e. at temperatures above the transition to an
ordered state. One can obtain an estimate for the ordering
temperature $T_c$ by considering the condition for an instability to
develop in form of a zero-frequency pole in $\chi_{\rm
3d}(\omega,\vec{k})$ \cite{RPA,schulz}. This pole indicates that the
system is unstable with respect to developing collective modes. In the
case of an antiferromagnet on a cubic lattice the instability
signals the spontaneous breakdown of spin rotational symmetry via
the emergence of N\'eel order and the collective modes are the
corresponding Goldstone modes (spinwaves). For the case of a quasi-1D
antiferromagnet on a cubic lattice, the critical
temperature obtained in this way is comparable to the experimentally
measured N\'eel temperature for ${\rm KCuF_3}$ \cite{schulz,Tennant95b}.
The leading corrections in the inverse coordination number of the
lattice can be calculated as well \cite{irkhin}. In our case the
instability condition reads 
\be
\label{instability_condition}
2J(k,k_y,k_z)\, \chi^{\alpha\alpha}(0,k)\Bigr|_{T=T_c}=1 \, ,
\ee
where $\alpha=x,z$. We find that for $\alpha=z$ the condition
(\ref{instability_condition}) leads to a weaker instability, i.e. occurring
at a lower temperature, except in the case of zero exchange
anisotropy. We therefore concentrate on $\alpha=x$ from now on.  
The instability will develop at the maximum of
$J(k,k_y,k_z)\, \chi^{\rm xx}(0,k)$. The maximum of the susceptibility of
a single chain $\chi^{\rm xx}(0,k)$ occurs at $k=\pi$, but because
$J(k,k/2,0)=0$ vanishes at this point the maximum of $J\chi^{\rm xx}$
will be shifted away from the antiferromagnetic wave number. 

Extremizing $J(k,k_y,k_z)\, \chi^{\rm xx}(0,k)$ with respect
to $k_y$ and $k_z$ yields $k_y=k/2 \text{ mod } (2\pi)$ and
$k^z=\pi$. Setting $k_y=k/2$, $k_z=\pi$ we find that the maximum of the
resulting expression occurs at a value $k=\pi+k_0$ such that
\bea
\label{extremum}
&&\frac{\pi\sinh (uk_0/2T_c)}{\cosh(uk_0/2T_c)-\cos(\frac{\pi\eta}{2})}
+\frac{2 \pi T_c}{u} \frac{\cos(k_0/2)}{J_z/J_\perp+2\sin(k_0/2)}\nn
&&-2\,{\rm Im}\, \Psi \( \frac{\eta}{4}+i\frac{u\,k_0}{4\pi T_c}\)\,=\,0\, ,
\eea
where $\Psi(x)$ is the digamma function. Equations (\ref{extremum}) and
(\ref{instability_condition}) with $k=\pi+k_0$, $k_y=(\pi+k_0)/2$, $k_z=\pi$
constitute two equations for the two unknowns $k_0$ and $T_c$. 
Let us emphasize that a solution with $k_0\neq 0$ corresponds to
an instability at an incommensurate wave number $\pi+k_0$. This
suggests that the resulting order occurs along the chain
direction with characteristic wave number $\pi+k_0$. 

In
general (\ref{instability_condition}) and (\ref{extremum}) can only be
solved numerically, which is easily done. An exception is the special
case $J_z=0$ and $\Delta=1$ (isotropic point), for which we can derive
analytic expressions. 

\subsubsection{Isotropic point ($\eta=1$)}

In the absence of an exchange anisotropy equation (\ref{extremum})
simplifies further  
\be
\frac{2 \pi T_c}{uk_0} + \pi \tanh (uk_0/2T_c)-2\, {\rm Im}\, \Psi \(
\frac{1}{4}+i\frac{uk_0}{4\pi T_c}\)\,=\,0\, .
\ee
Solving for $x_0=k_0/(4\pi T_c)$, we find
\be
x_0=\frac{\hbar u}{a_0}\frac{k_0}{ 4\pi T_c}\approx\pm 0.31 \ ,
\ee
where we have restored units ($k_0$ is measured in units of the
lattice spacing $a_0$ and $T$ in units of $k_B$). We note that
$k_0\neq 0$, which means that the instability develops at an {\sl
incommensurate} wave number close to $\pi$.

The remaining equation (\ref{instability_condition}) then becomes
\bea
&& 2J(k_0,0,0)\chi(0,k_0) =  -2 (J_z+2 J_{\perp}\sin(k_0/2))\chi(0,k_0) \nn
&&= \frac{1}{(2\pi)^{3/2}}
\left|\frac{\Gamma\left(\frac{1}{4}+ix_0\right)}
{\Gamma\left(\frac{3}{4}+ix_0\right)}\right|^2 
\( \frac{J_z}{T_c}+ \frac{4\pi x_0J_\perp a_0}{\hbar u} \)
\sqrt{\ln(\Lambda/T_c)} \nn
&& \approx  0.25 \(
\frac{J_z}{T_c}+8\,x_0\,\frac{J_{\perp}}{J_{\parallel}}\)\sqrt{\ln(\Lambda/T_c)}
= 1\, .
\eea
Here we have used the result for the spin velocity of the isotropic
Heisenberg spin 1/2 chain $u=\pi J_{\parallel} a_0/2$.

Let us first ignore the effects of the coupling $J_z$ between
planes. This case would correspond to a situation where the coupling
in $z$ direction is also frustrated, but much smaller than $J_\perp$.
The critical temperature and ordering wave number are then given by
\bea
\label{critical_temperature}
T_c&\approx& \Lambda\exp\Bigl(-2.60\,(J_{\parallel}/J_\perp)^2\Bigr)\ ,\\
|k_0| &\approx& 2.48 \, \frac{\Lambda}{J_{\parallel}}\, \exp \( -2.60\,
\(J_{\parallel}/J_\perp \)^2 \)\ . 
\label{k0}
\eea

These clearly show that the instability due to the frustrated
interchain coupling is {\sl extremely weak}. Firstly, the critical
temperature is orders of magnitude lower than its counterpart for the
square lattice. This is an indication how weak the tendency
to order is on a frustrated lattice. Even for a large
interchain coupling $J_{\parallel}/J_\perp = 0.4$, the critical temperature
is only $T_c \approx 9.10^{-6}$ K. The ordering wave number is
similarly small. Our results should be compared to the
zero-temperature results for the ordering wave number of 
\cite{Weihong99} and of \cite{Chung00}. The authors of
\cite{Weihong99} used linked-cluster expansion methods to
study ground state properties and excited states at zero temperature
of a rather general frustrated model, which contains the anisotropic
triangular lattice as a special case. They obtain  numerically (in
their notations our antiferromagnetic wave number corresponds to
$\pi/2$) 
\begin{equation}
q\simeq {\rm arccos}\Bigl(\frac{1}{4-6J_\parallel/J_\perp}\Bigr)\ ,
\end{equation}
which yields a much larger incommensuration than our result. In
\cite{Chung00} a Heisenberg model on an anisotropic triangular lattice
was studied by enlarging the spin-rotational symmetry group from SU(2)
to Sp(N) and then carrying out a large-N expansion.
It was concluded that for weak $J_\perp<0.125 J_\parallel$ there is no
incommensuration at all.  Given the approximate nature of our results,
they are certainly compatible with such a scenario.

In our approach the instability in the SU(2) invariant case is
actually caused by the marginally irrelevant current-current
interaction that gives rise to logarithmic corrections. The weakness
of the instability is due to the fact that it is caused by precisely
these logarithmic corrections.

\subsubsection{General case}
In the general case $J_z\neq0$ it is not possible to obtain simple
analytic expressions like (\ref{critical_temperature}) and (\ref{k0})
for $T_c$ and $k_0$ respectively and we have to resort to a numerical
solution of (\ref{instability_condition}) and (\ref{extremum}). We
present the results for various values of $J_z$ and $J_\perp$ in Table
1. 

\begin{center}
\begin{tabular}{|c||c|c|c|}
\hline
$J_z/J_\parallel$
& ${J_\perp}/{J_\parallel}$ =0.1
& ${J_\perp}/{J_\parallel}$ =0.2
& ${J_\perp}/{J_\parallel}$ =0.3\\
\hline\hline
0
&  $t_c=2.9 \ 10^{-112}$
&  $t_c=1.4 \ 10^{-27}$
&  $t_c=6.9 \ 10^{-11}$\\
&  $k_0=7.3 \ 10^{-112}$   
&  $k_0=3.6 \ 10^{-27}$
&  $k_0=1.7 \ 10^{-10}$\\
\hline
0.01
& $t_c=0.015$
& $t_c=0.017$
& $t_c=0.019$\\
&  $k_0=0.005$   
&  $k_0=0.011$
&  $k_0=0.019$\\
\hline
0.02
& $t_c=0.029$
& $t_c=0.032$
& $t_c=0.036$\\
&  $k_0=0.009$   
&  $k_0=0.020$
&  $k_0=0.034$\\
\hline
0.04
& $t_c=0.056$
& $t_c=0.060$
& $t_c=0.066$\\
&  $k_0=0.017$   
&  $k_0=0.036$
&  $k_0=0.060$\\
\hline
0.1
& $t_c=0.13$
& $t_c=0.14$
& $t_c=0.15$\\
&  $k_0=0.036$   
&  $k_0=0.076$
&  $k_0=0.13
$\\
\hline \hline
\end{tabular}
\end{center}
\underline{Table 1:} {\small Transition temperatures
$t_c=T_c/J_\parallel$ and ordering wave numbers $k_0$ for isotropic
exchange interaction and various values of the frustrated ($J_\perp$)
and antiferromagnetic ($J^z$) interchain couplings. }
\vskip .5cm

We see that that increasing $J_z$ leads to a significant increase of
the transition temperature {\sl and} the ordering wave number, whereas
increasing the frustrated coupling $J_\perp$ mainly increases the
value of $k_0$. 
The interesting observation is that a purely antiferromagnetic
coupling between the chains leads not only to a higher transition
temperature (as expected), but also to a significantly larger value of
the ordering wave number. If we take values similar to those for ${\rm
Cs_2CuCl_4}$  ($J_\perp/J\approx 0.33$, $J_z/J_\parallel\approx 0.05$)
we obtain a transition temperature of $t_c=0.084$ and an
incommensurablity of $k_0=0.083$. If we take subleading corrections in
the RG improvement into account \cite{Barzykin00a} these values
increase to $t_c=0.108$ and $k_0=0.104$ respectively.

These values (which should be taken with a degree of caution as
$J_\perp$ is not small) are roughly comparable to what is observed
for ${\rm Cs_2CuCl_4}$ \cite{Coldea00}, where $t_c\approx 0.145$ and
$k_0\approx 0.186$. 
As is shown in section \ref{sec:3dDM}, the presence of a DM interaction 
can further increase $t_c$ and $k_0$.

\hfill We may also study the effects of an exchange anisotropy $\Delta< 1$
along the z-direction is spin space (the chain Hamiltonian then is
given by(\ref{hxxz})). The analysis is completely analogous the
isotropic case. We find that the anisotropy enhances both the
transition temperature and the ordering wave number. For example, for
$\Delta=0.8$, $J_z=0.01J_\parallel$ and $J_\perp=0.1J_\parallel$ we
obtain $k_0=0.007$, $t_c=0.023$.
For $\Delta=0.5$, $J_z=0.01J_\parallel$ and $J_\perp=0.1J_\parallel$ we
have $k_0=0.013$, $t_c=0.033$.

%%%%%%%%%%%%%%%%%%%%%%%%%%%%%%%%%%%%%%%%%%%%%%%%%%%%%%%%%%%%%%%%%%%%%%%%%%%
\section{The effect of a uniform magnetic field}
%%%%%%%%%%%%%%%%%%%%%%%%%%%%%%%%%%%%%%%%%%%%%%%%%%%%%%%%%%%%%%%%%%%%%%%%%%%
\label{sec:H}

As we have mentioned in the introduction, the inspiration for the
present work are the observed properties of
Cs$_2$CuCl$_4$. The effects of an applied magnetic field
\cite{Coldea96,Coldea00} are particularly intriguing.
Let us therefore turn to the study of the effects of a uniform field
on the dynamical susceptibility, ordering temperature and ordering
wave number.
To this end, we must modify the dynamical susceptibility of a single
chain by taking into account the effect of the field. This can  be
done by utilizing exact results obtained by the Bethe Ansatz
\cite{vladb} (see the appendix of \cite{fab} for a brief summary).
For simplicity we consider the case of a single, isotropic Heisenberg
XXX chain ($\Delta=1$). In zero field we have full spin rotational
symmetry and transverse and longitudinal susceptibilites
coincide. There are gapless, transverse and longitudinal excitations
at $k=\pi$ and the spectral weight of the dynamical structure factor
is concentrated in the region around $k=\pi$. Application of a
magnetic field breaks the symmetry between transverse and longitudinal
susceptibilities. 

\begin{figure}[h]
\epsfxsize=3.5cm
(a)\epsfbox{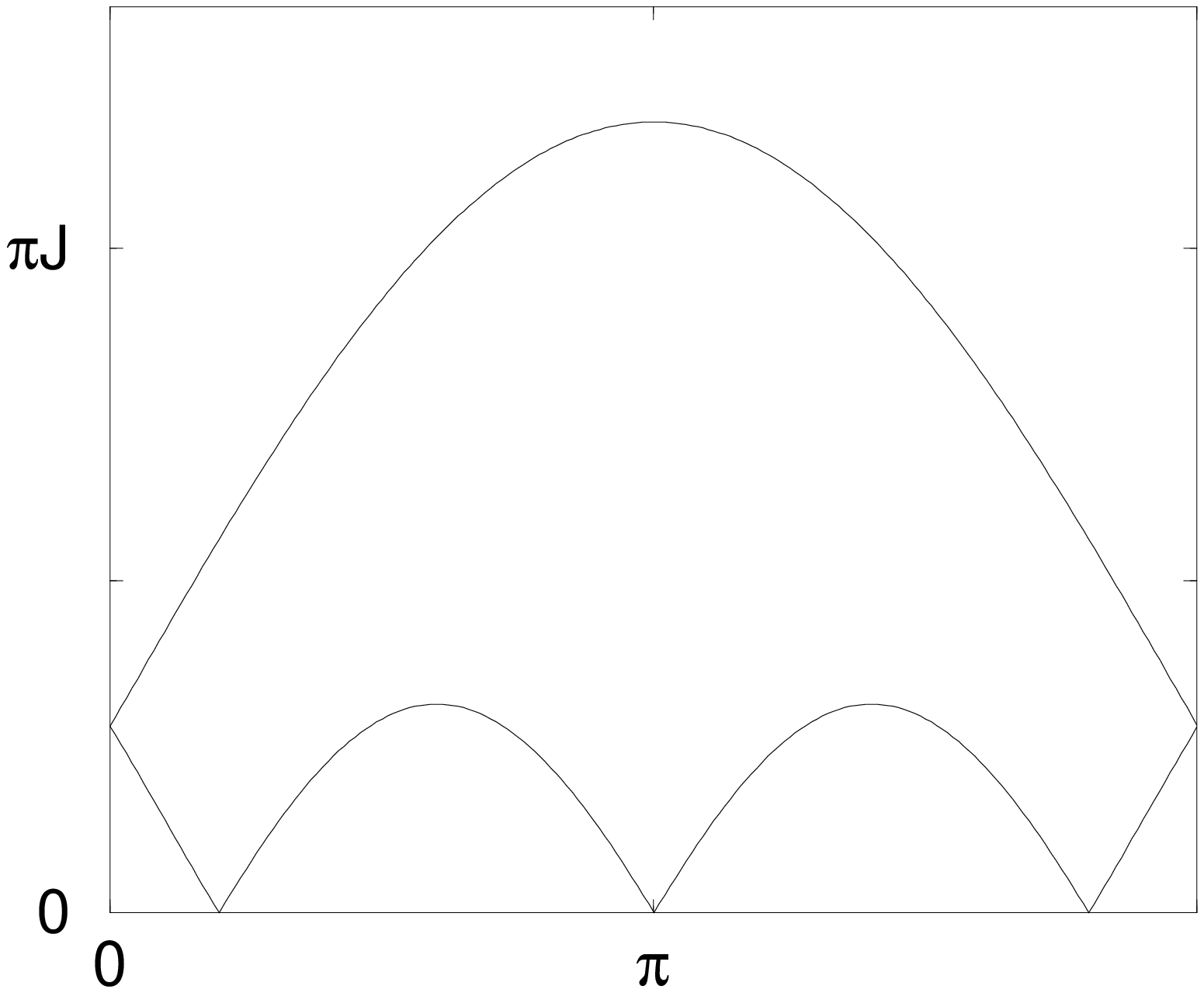}
\epsfxsize=3.5cm
(b)\epsfbox{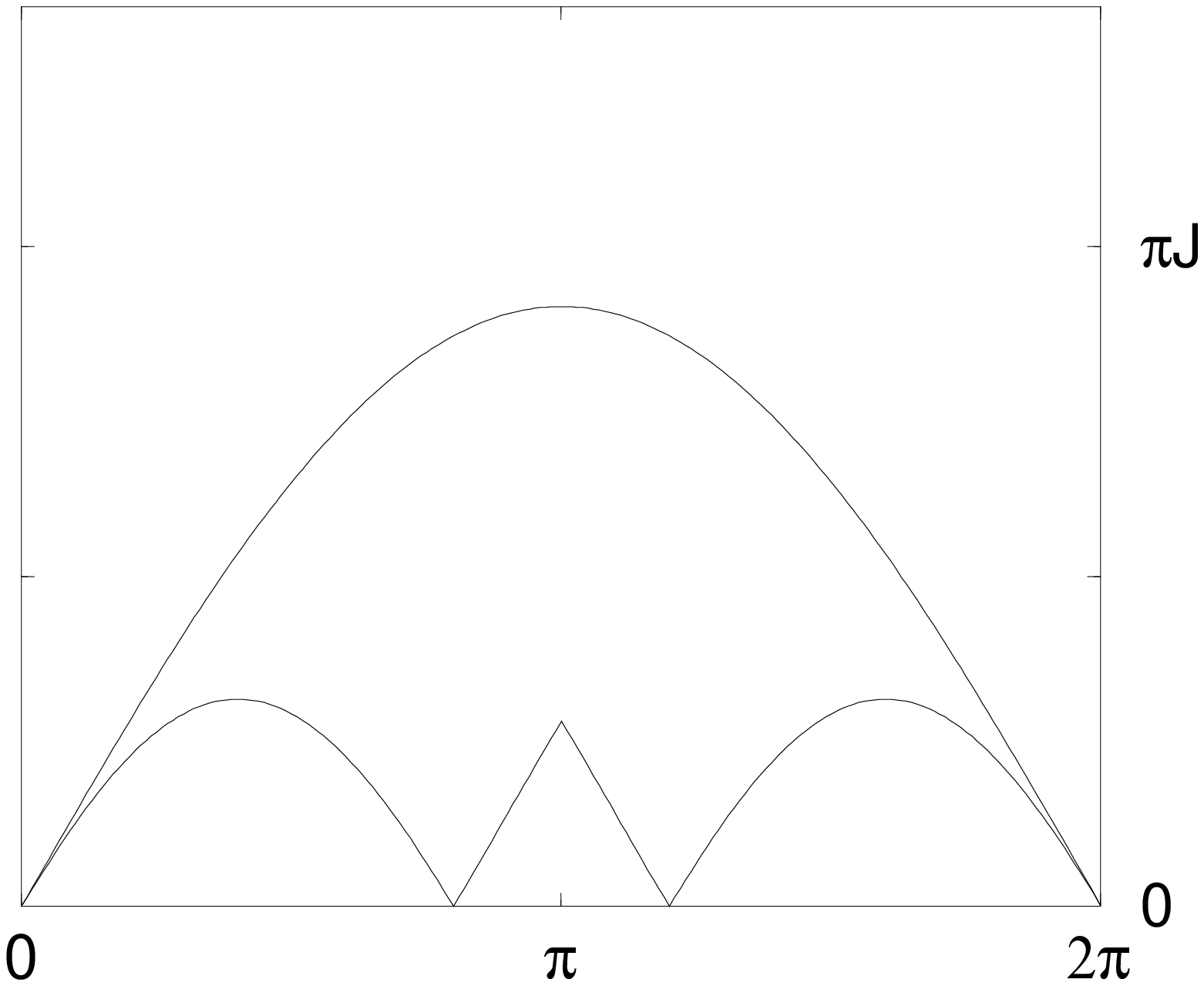}
\caption{Two-spinon continuum for the isotropic Heisenberg chain in a 
magnetic field (a) States with $\Delta S^z=1$ (b) States with $\Delta
S^z=0$. }  
\label{fig:dispH}
\end{figure}

The transverse susceptibility at low energies is dominated by the
contribution from the gapless states at $k=\pi$ in
Fig.\ref{fig:dispH}(a), whereas the most important contributions to
the longitudinal susceptibility at low energies comes from the gapless
states at the incommensurate wave numbers $\pi\pm\delta(H)$ in
Fig.\ref{fig:dispH}(b).

At $T=0$ the large-distance asymptotics of the transverse spin-spin
correlation functions in real space is
\bea
\langle S^x(\tau,y)S^x(0,0)\rangle&\simeq&
A_x(H) \frac{(-1)^{y/a_0}}{|y+iu(H)\tau|^{\eta(H)}}\ .
\label{realspace}
\eea
The magnetic field dependent spin velocity $u(H)$ and exponent
$\eta(H)$ can be calculated to high precision from the
Bethe Ansatz by solving certain linear integral equations
\cite{vladb}. 
The amplitude $A_x(H)$ is not known analytically but has very recently
been calculated numerically for several values of $H$ using a density
matrix renormalization group algorithm \cite{Furusaki00}. 

The asymptotic behaviour \r{realspace} holds for distances $|\tau+i
y/u(H)|$ large compared to $1/H$. The asymptotic behaviour of the
spin-spin correlator at very large distances is changed by the
application of a magnetic field, but at ``intermediate'' distances the
field does not play the same role. In order to utilize \r{realspace} we
therefore need sufficiently large magnetic fields. This is the case we
are interested in. For small fields \r{realspace} holds only at
extremely large distances and it is better to take the field into
account in RG improved perturbation theory. 

The dynamical susceptibility can now be determined by Fourier
transforming \r{realspace} and then analytically continuing the result
to real frequencies. The calculation is analogous to the one in zero
field \cite{schulz} and one obtains the following result
for the low-energy transverse susceptibility of a single,
isotropic Heisenberg XXX chain ($\Delta=1$) in a magnetic field
\bea
\chi^{\rm xx}(\omega,\pi - k,H)&=&\Phi(T,H)\frac{
\Gamma\(\frac{\eta(H)}{4}-i\frac{\omega-u(H)k}{4\pi
T}\)}{\Gamma\(1-\frac{\eta(H)}{4}-i\frac{\omega-u(H)k}{4\pi
T}\)}\nn
&&\times\frac{\Gamma\(\frac{\eta(H)}{4}-i\frac{\omega+u(H)k}{4\pi
T}\)}{\Gamma\(1-\frac{\eta(H)}{4}-i\frac{\omega+u(H)k}{4\pi T}\)} \, ,
\label{chiH}
\eea
where $|k| \ll 1$ and $\Phi(T,H)$ is given by
\bea
\Phi(T,H)&=&-A_x(H)
\frac{\sin\(\pi\eta(H)/2\)\Gamma^2\(1-\eta(H)/2\)}
{u(H)}\nn
&&\times \bigg[\frac{u(H)}{2 \pi T}\bigg]^{2-\eta(H)}.
\eea

The presently available results for the unknowns in
\r{realspace}, (\ref{chiH}) are summarized in Table 2.

\begin{center}
\begin{tabular}{|c|c|c|c|c|}
\hline
M(H) & $H/J_\parallel$ & $A_x(H)$ & $\eta(H)$ & $u(H)/J_\parallel a_0$\\
\hline\hline
0.05	&0.422	&0.121	&0.837	&1.501\\
\hline
0.10	&0.792	&0.120	&0.782	&1.398\\
\hline
0.15	&1.109	&0.118	&0.735	&1.259\\
\hline
0.20	&1.373	&0.117	&0.692	&1.093\\
\hline
0.25	&1.585	&0.112	&0.653	&0.911\\
\hline
0.30	&1.748	&0.106	&0.617	&0.721\\
\hline
0.35	&1.866	&0.095	&0.584	&0.529\\
\hline
0.40	&1.944	&0.081	&0.554	&0.342\\
\hline
0.45	&1.987	&0.061	&0.526	&0.165\\
\hline 
\end{tabular}
\end{center}
\underline{Table 2:} {\small 
Magnetization $M(H)$, spin velocity $u(H)$, exponent $\eta(H)$ and
amplitude $A_x(H)$ for different values of the uniform magnetic field $H$.
}
\vskip .5cm

As noted above, the result (\ref{chiH}) is least reliable at low fields
as it is obtained by Fourier transforming the large-distance
asymptotics of the real space correlation function. 

Using the results summarized in Table 2 in (\ref{chiH}) we obtain the
transverse susceptibility for a single chain. This result can now
again be combined with an RPA treatment of the interchain
couplings to obtain an expression for the susceptibility of the
quasi-1D. Following the same steps as in the absence of
a field we can determine a critical temperature and ordering wave
number.

\begin{figure}[h]
\epsfxsize=8cm
\epsfbox{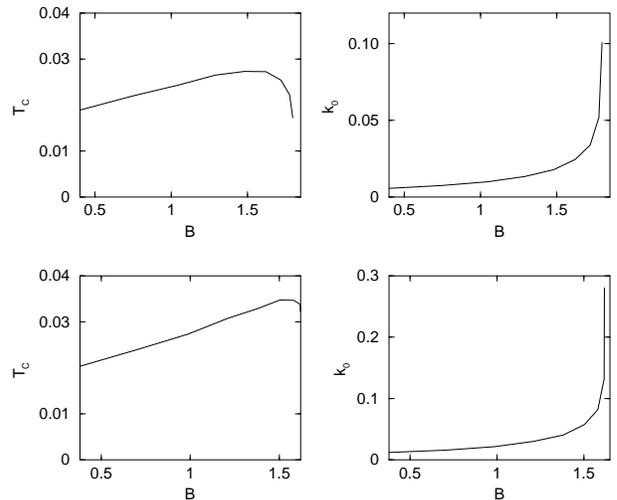}
\caption{Upper two graphs: critical temperature $T_c$ and ordering wave
number $k_0$ extracted from the transverse susceptibility as functions
of the applied field $B=H/J_\parallel$ for isotropic Heisenberg chains
coupled by a frustrated in-plane coupling $J_\perp=0.1J_\parallel$ and
antiferromagnetic inter-plane coupling $J_z=0.01J_\parallel$.
Lower two graphs: the same for $J_\perp=0.2 J_\parallel$.
} 
\label{fig:magn01}
\end{figure}

In Fig.\ref{fig:magn01} we plot the ordering temperature and wave
number as a function of the applied field. We see that at first both
increase with the field, but $T_c(H)$ eventually goes through a
maximum and then decreases. This is in qualitative agreement with what
has been observed for ${\rm Cs_2CuCl_4}$. If we switch on an exchange
anisotropy along the z-direction in spin space the qualitative
behaviour of the $T_c$ and $k_0$ as functions of $H$ remain the same.

So far we have only considered the transverse susceptibility. In the
presence of a field we still have to check for possible instabilities
in the longitudinal susceptibility as the maxima of
$\chi^{\rm xx}(0,k,H)$ and $\chi^{\rm zz}(0,k,H)$ occur for different values
of $k$. It is important to note that the physical nature of a
longitudinal instability is very different from that of a transverse
one. The latter is associated with the spontaneous breakdown
of the spin rotational symmetry around the direction of the magnetic
field, whereas a longitudinal instability does not break any
continuous symmetries, but rather corresponds to the formation of a
spin-density wave in the ground state.

The longitudinal susceptibility in a field can be calculated along the
same lines as outlined above for the transverse ones. However, the
longitudinal real-space correlation function decays {\sl faster} with
distance as the field is increased. Therefore it is likely that
Fourier transforming the large-distance asymptotics leads to a less
reliable result than for the transverse susceptibilities. It would be
very interesting to investigate this issue by numerical methods.
Fourier transformation of the large-distance asymptotics and
analytical continuation yields
\bea
&&\chi^{\rm zz}(\omega,\pi \pm \delta - k,H)=\Phi^\prime(T,H)\nn
&&\times
\frac{\Gamma\(\frac{1}{4\eta(H)}-i\frac{\omega-u(H)k}{4\pi T}\)}
{\Gamma\(1-\frac{1}{4\eta(H)}-i\frac{\omega-u(H)k}{4\pi T}\)}
\frac{\Gamma\(\frac{1}{4\eta(H)}-i\frac{\omega+u(H)k}{4\pi T}\)}
{\Gamma\(1-\frac{1}{4\eta(H)}-i\frac{\omega+u(H)k}{4\pi T}\)},\nn
\label{chihzz}
\eea
where $\delta=2\pi M(H)$ and $\Phi^\prime(T,H)$ is given by
\bea
\Phi^\prime(T,H)&=&
-A_z(H) \frac{\sin(\pi/2\eta(H))\Gamma^2(1-1/2\eta(H))}{u(H)}\nn
&&\quad\times
\bigg[\frac{u(H)}{2\pi T}\bigg]^{2-1/\eta(H)}.
\eea 
The amplitude $A_z(H)$ is again known numerically \cite{Furusaki00},
so that we can repeat the analysis carried out above for the
transverse susceptibility. 
We find that in general an instability in the longitudinal
susceptibility exists at a critical temperature comparable to the one
for the transverse instability. The corresponding characteristic wave
number $k_0$ is located in the vicinity of $\pi\pm\delta=\pi\pm 2\pi
M(H)$, where $M(H)$ is the magnetization of uncoupled chains. Note that
$\pi\pm\delta$ are the points where incommensurable soft modes
exist in the $\Delta S^z=0$ sector of the excitation spectrum of
individual chains (see Fig.\ref{fig:dispH}(b)).
For large values of the applied field $H$, $\pi\pm\delta$ approach $0$
and $2\pi$ respectively. We note that in this regime the further
complication arises that it is no longer appropriate to neglect the
contributions (as has been done in (\ref{chihzz})) of the smooth,
non-oscillating terms in the spin-spin correlation functions. 

As a function of the applied field $H$, the transition temperature of
the longitudinal instability $t^{\rm long}_c(H)$ first increases, goes
through a maximum and then decreases again. This is similar to what we
obtained for the transverse instability. However, in contrast to the
transverse instability the maximum of $t^{\rm long}_c(H)$ occurs at
small magnetization. 

Which instability is dominant is rather sensitive to the values of
$J_\perp$, $J_z$ and the applied field $H$. For
$J_\perp=0.1J_\parallel$, $J_z=0.01J_\parallel$ we find that the
transverse instability occurs at a higher temperature except for small
fields, where the longitudinal susceptibility appears to dominate. On
the other hand, as we have pointed out above, for small fields our
results for the Fourier transforms are least reliable.

So far we have worked with a spin rotationally symmetric Hamiltonian.
Let us now consider the effects of an exchange anisotropy. We have to
distinguish two cases, depending whether or not the applied field is
along the direction of the exchange anisotropy. In the former case the
analysis is completely analogous to the spin rotationally symmetric
case and we find a behaviour $T_c(H)$ and $k_0(H)$ very similar to
Fig.\ref{fig:magn01}. The transition temperature increases with
increasing field, goes through a maximum and eventually decreases
again. 

Let us now turn to the case where we have an exchange anisotropy in
z-direction and apply the field along the x-direction. The effect of
the magnetic field is now to generate an excitation gap. This can be
seen as follows. The chain Hamiltonian is  
\bea
{\cal H}_{\rm XXZ,H}&=&J_{\parallel}\sum_j S^x_jS^x_{j+1}+S^y_jS^y_{j+1}+\Delta
S^z_jS^z_{j+1}\nn 
&&+H\sum_j S^x_j = {\cal H}_0+{\cal H}_1\ ,
\label{XXZH}
\eea
where ${\cal H}_0$ is the Hamiltonian of the anisotropic spin-1/2 chain
and 
\be
{\cal H}_1= H\sum_j S^x_j. 
\label{pert}
\ee
Let us study the effect of the perturbing operator by bosonizing
at the critical point defined by ${\cal H}_0$ and then perturbing this
fixed point theory by (\ref{pert}). The bosonized form of ${\cal H}_0$
is
\be
{\cal H}_0 = \frac{1}{2}\int dx[(\partial_x\Theta)^2 +
(\partial_x\Phi)^2]\,,
\ee
where $\Phi$ is a canonical bosonic field and $\Theta$ is the dual field. 
The uniform part of the $x$-component of the magnetization is
given by the product of two operators 
\be
H\,\cos\(\sqrt{2\pi\eta} \, \Theta\)\,\cos \(\sqrt{2\pi/\eta}\, \Phi \),
\ee
where $\eta$ has been defined in Eq.(\ref{eta}).
These  operators are formally relevant and not being Lorentz scalars
they belong to the class of perturbations with non-zero conformal
spin. Such operators require  a special treatment 
(see \cite{GNT}, chapters 8 and 20 and references therein). 
In particular, to second order, the following two spin-zero fields are
generated 
\bea
 \cos(\sqrt{8\pi\eta}\, \Theta), ~~ \cos(\sqrt{8\pi/\eta}\,\Phi).
\eea
The resulting problem has been considered in detail in Chapter 20 of \cite{GNT}.
Because we have $\eta < 1$, the renormalization group equations flow to a Sine-Gordon model
in the dual field. For weak fields $H$, a spectral gap of order of 
\be
M \propto H
\ee
is generated. The gap grows with increasing magnetic field. Clearly,
the growth of the gap will lead to a decrease in the transition
temperature $T_c$ - the effect observed in ${\rm Cs_2CuCl_4}$ when the
field is applied along the $c$ direction. 

There is a second way to see that application of a uniform field at an
angle to the exchange anisotropy induces a spectral gap
\cite{shura}. If we consider the Hamitonian
\bea
{\cal H}_{\rm ZXX,H}&=&J_{\parallel}\sum_j
S^y_jS^y_{j+1}+S^z_jS^z_{j+1}+\Delta S^x_jS^x_{j+1}\nn 
&&+H\sum_j S^z_j
\eea
and bosonize the isotropic Heisenberg chain in a field first and then
take the exchange anisotropy into account as a perturbation, we obtain
a Sine-Gordon model for the dual field \cite{shura,fab}. The cosine
term in the Sine-Gordon model is relevant and generates a spectral
gap. The spectrum of the Sine-Gordon model consists of soliton and
antisoliton only and the dynamical structure factor (in the xy plane) 
displays an incoherent two-particle continuum. This suggests that the
elementary excitations are {\sl massive spinons}.

%%%%%%%%%%%%%%%%%%%%%%%%%%%%%%%%%%%%%%%%%%%%%%%%%%%%%%%%%%%%%%%%%%%%%%
\section{A single spin-1/2 XXX chain with DM interaction} 
%%%%%%%%%%%%%%%%%%%%%%%%%%%%%%%%%%%%%%%%%%%%%%%%%%%%%%%%%%%%%%%%%%%%%%
\label{sec:1dDM}

\begin{figure}[h]
\epsfxsize=8cm
\epsfbox{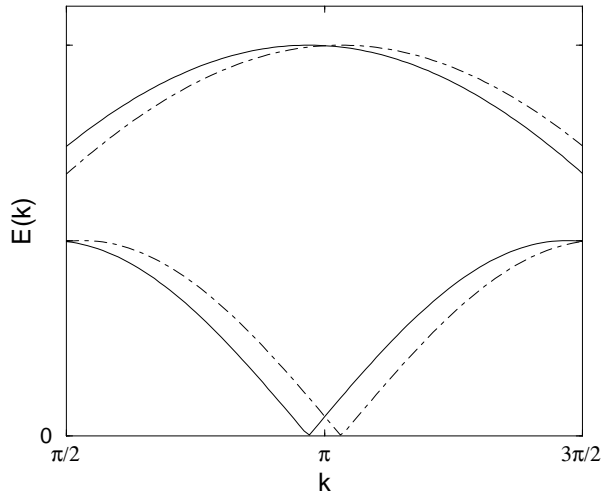}
\caption{
Schematic two-spinon dispersion in the vicinity of $k=\pi$ in the
sector $\Delta S^z=\pm 1$ for the isotropic Heisenberg chain with DM
interaction. } 
\label{fig:disp_dm}
\end{figure}

As discussed in \cite{Coldea97b}, DM interactions may play a role in
accounting for all the observed properties of Cs$_2$CuCl$_4$. We
therefore consider now a single, isotropic spin-1/2 Heisenberg chain
with a DM interaction along the $z$ direction in spin space
\begin{eqnarray}
\HDM &=&J^\prime\sum_{j=1}^L S^x_jS^x_{j+1}+S^y_jS^y_{j+1}+ S^z_jS^z_{j+1}\nn
&&+D\sum_j S^x_jS^y_{j+1}-S^y_jS^x_{j+1}\ .
\label{hdm}
\end{eqnarray}
It is well known that (\ref{hdm}) with periodic boundary conditions is
equivalent to an XXZ chain with twisted boundary conditions. Indeed, a
local rotation around the z-axis
\begin{eqnarray}
S^+_j&=& e^{-ij\theta}\ \tilde{S}^+_j, ~~ S^-_j = e^{ij\theta}\ \tilde{S}^-_j,\nn
S^z_j&=& \tilde{S}^z_j\ ,
\label{trafo}
\end{eqnarray}
with $\theta=-{\rm arctan}(D/J^\prime)$ maps the Hamiltonian (\ref{hdm}) onto
\begin{equation}
\HDM =J\sum_{j=1}^L
\tilde{S}^x_j\tilde{S}^x_{j+1}+\tilde{S}^y_j\tilde{S}^y_{j+1}+\Delta
\tilde{S}^z_j\tilde{S}^z_{j+1}. 
\end{equation}
Here $J={J^\prime}/{\cos\theta}$ and $\Delta=\cos\theta$. For a system
with open boundary conditions there are no further changes. In
particular, we can rest assured that bulk correlation functions in 
(\ref{hdm}) can be obtained from bulk correlators in the anisotropic
spin-1/2 Heisenberg chain. The excitation spectrum of a spin-1/2
Heisenberg chain with DM interaction can be obtained from that of the
corresponding XXZ chain by taking into account the shift in momentum
induced by the mapping (\ref{trafo}). In Fig.\ref{fig:disp_dm} we show
the qualitative excitation spectrum with quantum numbers
$\Delta S^z=\pm 1$ around momentum $\pi$. There are incommensurate soft
modes at $\pi\pm\theta$. Excitations with $\Delta S^z=0$ stay
commensurate i.e. become soft at momentum $\pi$.

Using the above mapping we can now express bulk correlation functions 
of the spin-1/2 chain with DM interaction in terms of correlation
functions of an XXZ chain with exchange $J/\cos\theta$ and anisotropy
$\Delta=\cos\theta$. For example, 
\begin{eqnarray}
\langle S^+_jS^-_{j+k}\rangle_{\text{\tiny DM}}&=& e^{ik\theta}\ \langle
\tilde{S}^+_j\tilde{S}^-_{j+k}\rangle_{\text{\tiny XXZ}}, \nn
\langle S^-_jS^+_{j+k}\rangle_{\text{\tiny DM}}&=& e^{-ik\theta}\ \langle
\tilde{S}^-_j\tilde{S}^+_{j+k}\rangle_{\text{\tiny XXZ}},
\end{eqnarray}
where we have used that by global spin rotational symmetry (of the
bulk) around the $z$-axis 
\begin{equation}
\langle \tilde{S}^x_j\tilde{S}^x_{j+k}\rangle_{\text{ \tiny XXZ}} =\langle 
\tilde{S}^y_j\tilde{S}^y_{j+k}\rangle_{\text{\tiny XXZ}}\ ,\quad
\langle \tilde{S}^x_j\tilde{S}^y_{j+k}\rangle_{\text{\tiny XXZ}}=0.
\end{equation}
This allows us to express the dynamical magnetic susceptibility of the
model (\ref{hdm}) in terms of the results for the Heisenberg XXZ chain
\bea
\chi^{+-}_{\text{ \tiny DM}}(\omega,k)&=&
\chi^{+-}(\omega,k-\theta),\nn
\chi^{-+}_{\text{\tiny DM}}(\omega,k) &=&  
 \chi^{-+}(\omega,k+\theta), \nn
\chi^{\rm zz}_{\text{ \tiny DM}}(\omega,k)&=&\chi^{\rm zz}(\omega,k).
\eea

\begin{figure}[h]
\epsfxsize=8cm 
\epsfbox{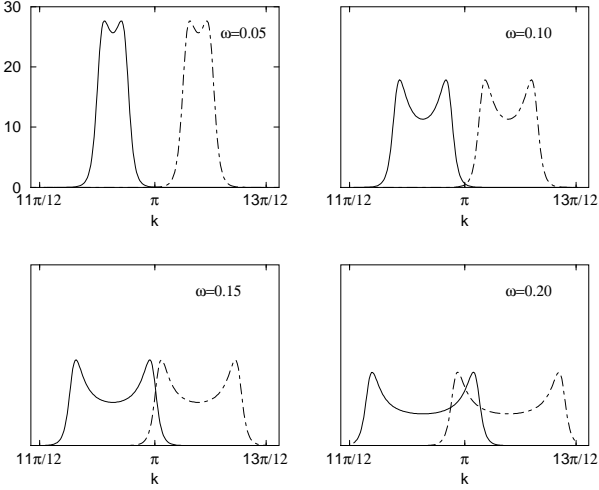} 
\caption{Imaginary part of the dynamical susceptibility $-{\rm
Im}\chi_{\text{\tiny DM}}^{+-}(\omega,k)$ (full line) and 
$-{\rm Im}\chi_{\text{\tiny DM}}^{-+}(\omega,k)$ (dashed-dotted line)
as functions of $q$ for four
different values of $\omega/J$ at a temperature of
$k_BT/J=0.01$. The DM angle is chosen to be $\theta=0.1$.
} 
\label{fig:chi1d_dm}
\end{figure}

In Fig.\ref{fig:chi1d_dm} we plot $-{\rm Im}\chi^{\pm\mp}_{\text{\tiny  DM}}(\omega,k)$
for a DM angle of $\theta=0.1$ as a function of $q$ for four different
values of $\omega/J_\parallel$ at a temperature of $k_BT/J_{\parallel}=0.01$. 
Had we plotted instead
\be
\chi^{\rm xx}(\omega,k)={1\over 4}\( \chi^{+-}(\omega, k-\theta)+
\chi^{-+}(\omega, k+\theta) \) \, ,
\ee
we would have seen an incommensurate four-peak structure.
Because of the chirality introduced by the interaction,
$\chi^{+-}_{\text{\tiny  DM}}$ and $\chi^{-+}_{\text{\tiny  DM}}$ now
differ ($\vec{u}_z.\vec{S}(\vec{k})
\times \vec{S}(-\vec{k})$ has a nonvanishing expectation value).

%%%%%%%%%%%%%%%%%%%%%%%%%%%%%%%%%%%%%%%%%%%%%%%%%%%%%%%%%%%%%%%%%%%%%%%
\section{Dynamical susceptibility for coupled Heisenberg chains with
DM interaction}
%%%%%%%%%%%%%%%%%%%%%%%%%%%%%%%%%%%%%%%%%%%%%%%%%%%%%%%%%%%%%%%%%%%%%%%
\label{sec:3dDM}

In the presence of a DM interaction along the z-direction in spin
space, the RPA result for the dynamical susceptibility of coupled
chains is given by
\begin{eqnarray}
\chi_{\rm 3d}^{+-}(\omega,\vec{k})&=&
\frac{\chi^{+-}_{\rm}(\omega,k-\theta)}  
{1-J(k,k_y,k_z)\,\chi^{+-}_{\rm }(\omega,k-\theta)}\ ,\nn
\chi_{\rm 3d}^{-+}(\omega,\vec{k})&=&
\chi_{\rm 3d}^{+-}(\omega,\vec{k}, \theta \rightarrow -\theta)\ ,\nn
\chi_{\rm 3d}^{\rm zz}(\omega,\vec{k})&=& 
\frac{\chi^{zz}(\omega,k)}
{1-2\,J(k,k_y,k_z)\,\chi^{zz}(\omega,k)}\ ,
\label{chi3dDM}
\end{eqnarray}
where $J(k,k_y,k_z)$ is given by (\ref{jofk}) and
$\chi^{\alpha\beta}(\omega,k)$ is the dynamical susceptibility of a
single Heisenberg chain. We note that the RPA expressions for
$\chi^{\rm \alpha\alpha}_{\rm 3d}(\omega,\vec{k})$, $\alpha=x,y$, are
not simply (\ref{chi3d}) with $\chi^{\alpha\alpha}(\omega,k)$ replaced
by $\chi_{\text{\tiny DM}}^{\alpha\alpha}(\omega,k)$. Instead,
$\chi^{\rm xx}_{\rm 3d}=\chi^{\rm yy}_{\rm 3d}$ and $\chi^{\rm xy}_{\rm 3d}=-\chi^{\rm yx}_{\rm3d}$
are obtained from $\chi^{\rm +-}_{\rm 3d}$ and $\chi^{\rm -+}_{\rm 3d}$ by
\be
\chi^{\rm xx}_{\rm 3d}=\frac{1}{4}\( \chi^{\rm +-}_{\rm 3d} + \chi^{\rm -+}_{\rm 3d} \)\,, \;
\chi^{\rm xy}_{\rm 3d}=\frac{1}{4i}\( \chi^{\rm +-}_{\rm 3d} - \chi^{\rm -+}_{\rm 3d} \) \, .
\ee

\begin{figure}[h]
\epsfxsize=8cm
\epsfbox{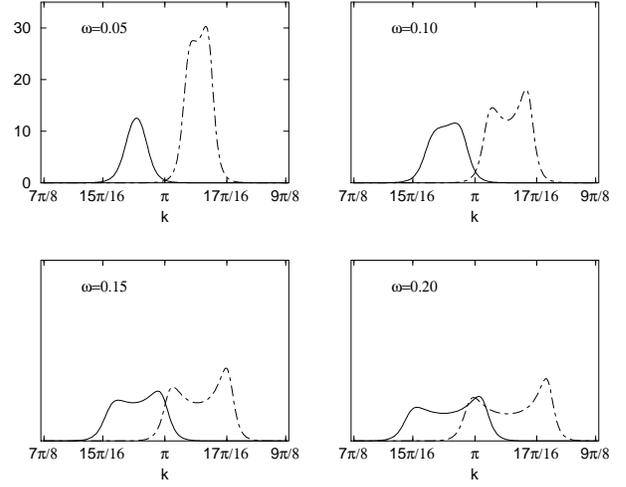}
\caption{
$-{\rm Im}\chi^{+-}_{\rm 3d}(\omega,\vec{k})$ (solid line)
and $-{\rm Im}\chi^{-+}_{\rm 3d}(\omega,\vec{k})$ (dashed-dotted line)
for a DM angle $\theta=0.1$, $J_\perp/J_{\parallel}=0.1$, 
$T/J_{\parallel}=0.02$ and $J^z=0$. The frustrated interchain coupling
breaks the symmetry around  $k=\pi$.} 
\label{fig:chixx_2d_dm}
\end{figure}

\begin{figure}[h]
\epsfxsize=8cm
\epsfbox{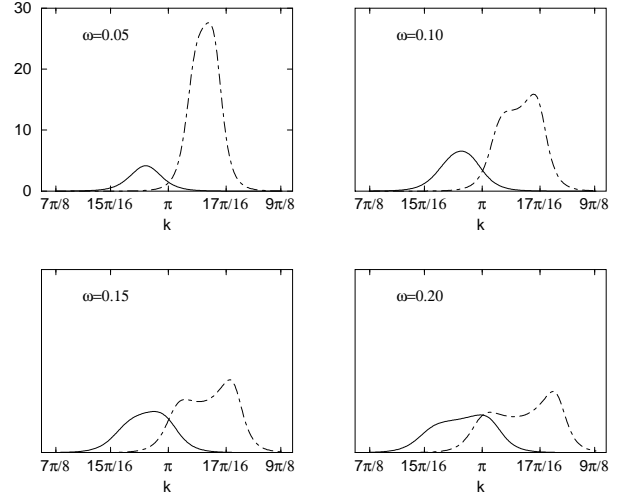}
\caption{
$-{\rm Im}\chi^{\rm +-}_{\rm 3d}(\omega,\vec{k})$ (solid line)
and $-{\rm Im}\chi^{\rm -+}_{\rm 3d}(\omega,\vec{k})$ (dashed-dotted line)
for a DM angle of $\theta=0.1$, $J_\perp/J_{\parallel}=0.2$, $T/J_{\parallel}=0.04$ and $J^z=0$.}
\label{fig:chixx_3d_dm}
\end{figure}

In Figs \ref{fig:chixx_2d_dm} and \ref{fig:chixx_3d_dm} we plot the
imaginary parts of the dynamical susceptibilities $\chi^{+-}_{\rm
3d}(\omega,k)$ $\chi^{-+}_{\rm DM}(\omega,k)$ as functions of the
momentum transfer along the chain direction for a DM angle of
$\theta=0.1$, $J_z=0$ (the behaviour for $J_z\neq 0$ is qualitatively
the same) and two different values of $J_\perp$. The effect of the
frustrated interchain coupling is again to remove the symmetry around
$k=\pi$. 
%For $J_\perp/J=0.2$
%the four-peak structure characteristic of uncoupled spin-1/2 chains
%with DM interaction has largely disappeared.
%For $J_\perp/J=0.2$, the contribution of $\chi^{+-}_{\rm 3d}$
%to the total weight of $\chi_{\rm 3d}^{\rm xx}$,
%which was the same as the contribution of $\chi^{-+}_{\rm 3d}$
%for the uncoupled spin-1/2 chains with DM interaction,
%has largely reduced.

Let us now determine the transition temperature and ordering wave number
in the presence of a DM interaction. We will see that the effect of the
DM interaction is to significantly increase both quantities.
The instability conditions read
\bea
\label{instability_condition_DM}
J(k,k_y,k_z)\, \chi^{+-}(0,k - \theta)=1 \ , \nn
J(k,k_y,k_z)\, \chi^{-+}(0,k + \theta)=1\ ,
\eea
where the 1D dynamical susceptibilities fulfil $\chi^{+-}(0,k) =
\chi^{-+}(0,k)$. 
As before the instability develops at the maximum of $J(k,k_y,k_z)
\chi^{+-}(0,k \pm \theta)$. In the absence of a DM
interaction ($\theta = 0$) this
maximum was shifted away from the maximum of the susceptibility for a
single chain at $k=\pi$, because $J(k,k/2,\pi)$ vanished precisely at
$k=\pi$. In the presence of a DM interaction the maxima of the
single-chain susceptibility occurs at $\pi\pm\theta$ and the effect of
the frustrated interchain coupling is therefore different. We find
that the numerical value of $k$ at which the instability develops is
largely determined by the DM interaction and to a lesser degree by the
frustration. However, the frustrated interchain coupling destroys the
symmetry in momentum space around $k=\pi$ and determines whether the
instability will develop in the vicinity of $\pi+\theta$ or of
$\pi-\theta$. 
The fact that the transition temperature is increased by the DM
interaction can be understood by considering the susceptibility of a
single chain with DM interaction. As we have shown in section
\ref{sec:1dDM} the corresponding Hamiltonian maps onto an an
anisotropic Heisenberg XXZ chain. It follows from (\ref{chioq}) that
for $\Delta<1$ the transverse susceptibility is enhanced as compared
to the isotropic case $\Delta=1$, which in turn leads to a higher $T_c$.

Solving (\ref{instability_condition_DM}) and the equation for the
maximum of $J(k,k_y,k_z)\, \chi^{+-}(0,k\pm\theta)$ numerically, we
obtain the results for the transition temperature $T_c$ and
ordering wave number $k_0$ shown in Table 3.

\begin{center}
\begin{tabular}{|c||c|c|c|}
\hline
$\theta$
& ${J_\perp}/{J_\parallel}$ =0.1
& ${J_\perp}/{J_\parallel}$ =0.2
& ${J_\perp}/{J_\parallel}$ =0.3\\
\hline\hline
0.05
& $t_c=0.008$
& $t_c=0.017$
& $t_c=0.028$\\
&  $k_0=0.053$   
&  $k_0=0.061$
&  $k_0=0.076$\\
\hline
0.10
& $t_c=0.016$
& $t_c=0.032$
& $t_c=0.051$\\
&  $k_0=0.10$   
&  $k_0=0.12$
&  $k_0=0.14$\\
\hline
0.15
& $t_c=0.023$
& $t_c=0.046$
& $t_c=0.073$\\
&  $k_0=0.16$   
&  $k_0=0.18$
&  $k_0=0.21$\\
\hline
0.20
& $t_c=0.031$
& $t_c=0.060$
& $t_c=0.095$\\
&  $k_0=0.21$   
&  $k_0=0.23$
&  $k_0=0.27$\\
\hline
\end{tabular}
\end{center}
\underline{Table 3:} {\small Transition temperatures
$t_c=T_c/J_\parallel$ and ordering wave numbers $k_0$ for various values
of the frustrated ($J_\perp$) interchain coupling and DM angle
$\theta$. The interplane coupling $J_z$ is taken to be zero. The
effect of a small $J_z$ on $T_c$, $k_0$ is negligible compared to the
effect of the DM interaction.
}
\vskip .5cm

We see that a strong DM interaction leads to much larger values for
$T_c$ and $k_0$ than a frustrated interchain coupling
alone. For the values of couplings observed in ${\rm Cs_2CuCl_4}$
$J_\perp=0.33 J$, $J_z\approx 0.05J$,  $D\approx 0.05 J$ we obtain
$t_c=0.11$, $k_0=0.14$, which are close to the measured
values. However, recent evidence suggests that the DM interaction in
${\rm Cs_2CuCl_4}$ is not of the kind considered in this section
\cite{radu}. 

%%%%%%%%%%%%%%%%%%%%%%%%%%%%%%%%%%%%%%%%%%%%%%%%%%%%%%%%%%%%%%%%%%%%
\section{DM interaction and a magnetic field}
\label{sec:HDM}
%%%%%%%%%%%%%%%%%%%%%%%%%%%%%%%%%%%%%%%%%%%%%%%%%%%%%%%%%%%%%%%%%%

\begin{figure}[h]
\epsfxsize=7.9cm
\epsfbox{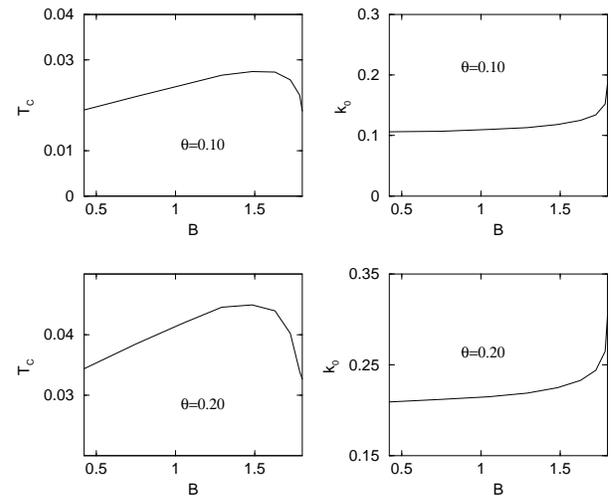}
\caption{Critical temperature $T_c$ and ordering wave number $k_0$ as
functions of the applied field $B=H/J_\parallel$ for isotropic
Heisenberg chains with a DM interaction, coupled by a frustrated
in-plane coupling $J_\perp=0.1J_\parallel$ and no antiferromagnetic
inter-plane coupling $J_z=0$. The DM angle is indicated in the figures.} 
\label{fig:magn02}
\end{figure}

Finally, let us investigate the case where we have both a DM
interaction and a magnetic field. Now it is crucial whether or not the
field is applied along the direction singled out by the DM interaction. 
If it is, the system remains gapless and we can proceed along the same
lines as before. If the magnetic field is applied at an angle to the
DM interaction, we believe that an excitation gap is generated \cite{BE}. Let us
consider the former case. The chain Hamiltonian is of the form
\bea
{\cal H}_{\text{\tiny DM,H}}&=&J^\prime\sum_{j=1}^L S^x_jS^x_{j+1}+S^y_jS^y_{j+1}+
S^z_jS^z_{j+1}\nn 
&&+D\sum_j S^x_jS^y_{j+1}-S^y_jS^x_{j+1}+H\sum_j S^z_j\ .
\label{hdmH}
\eea
By means of the unitary transformation (\ref{trafo}) this maps onto an
anisotropic Heisenberg XXZ chain in a field

\bea
{\cal H}_{\text{\tiny DM,H}}&=&J\sum_{j=1}^L
\tilde{S}^x_j\tilde{S}^x_{j+1}+\tilde{S}^y_j\tilde{S}^y_{j+1}+\Delta
\tilde{S}^z_j\tilde{S}^z_{j+1}\nn
&&+H\sum_j\tilde{S}^z_j\ ,
\label{DMH}
\eea
were $J={J^\prime}/{\cos\theta}$ and $\Delta=\cos\theta$
($\theta=-{\rm arctan}(D/J^\prime)$ as before).
The model (\ref{DMH}) remains gapless and we can determine the
finite-temperature dynamical susceptibility by the same methods we
used in the absence of a DM interaction in section \ref{sec:H} above.
The result for the transverse susceptibility is of the form
(\ref{chiH}), where the exponent $\eta(H,D)$ and velocity $u(H,D)$ can
again be determined from the Bethe Ansatz, whereas the amplitude
$A_x(H)$ is known numerically \cite{Furusaki00}.
Taking the interchain couplings into account in RPA and then looking
for instabilities as before, we obtain the results shown in
Fig.\ref{fig:magn02}. 

The behaviour of $T_c(H)$ and $k_0(H)$ as functions of the applied
field is similar to what we found in the absence of a DM
interaction. However, like for the zero field case a strong DM
interaction leads to a significant increase in the absolute values of
$T_c$ and $k_0$.

\section{General Dzyaloshinskii-Moriya interaction}
\label{sec:DMgeneral}

In this section we discuss how to treat more general types of DM
interactions within the framework of our coupled chain approach.

\subsection{Dzyaloshinskii-Morya interaction along the chains}
\label{ssec:DM1}
In a compound such as ${\rm Cs_2CuCl_4}$, the three dimensional
elementary cell contains four Cu$^{2+}$ ions \cite{structure}.
Two of them, say (1) and (2), lie within a planar layer and are
coupled by a relatively strong, frustrated exchange interaction. The
other two, say (3) and (4), lie in the layer above and are coupled
weakly to (1) and (2) \cite{Coldea97b}. 

In the case where it involves spins along the chain direction,
the most general form of the DM interaction which is allowed by
symmetries \cite{Coldea97b} is
\bea
{\cal I}_{\text{\tiny DM}} & =&  \sum_{n,m} \vec{D}_1 \cdot
\left[\vec{S}^{(1)}_{n,m} \times \vec{S}^{(1)}_{n+1,m} \right] \nn
&+&  \vec{D}_2 \cdot \left[ \vec{S}^{(2)}_{n,m+{1\over2}}
\times \vec{S}^{(2)}_{n+1,m+{1\over 2}} \right] \, .
\eea
Generically, this interaction requires the distinction between the sets (1) and (2).
(Up to now, we had only considered the case $\vec{D}_1=\vec{D}_2$).

As a consequence, and from now on, the Heisenberg chains forming up
the triangular planar lattice will not necessarily be considered all
equivalent, but belonging alternatively to sets (1) or (2). 
This doubles the primitive cell chosen in figure \ref{fig:fbz} in the
y-direction. The magnetic Hamiltonian that takes the difference
between the two types of spins into account is
\bea
\HDM &=& J_{\parallel}\sum_{m,n} \vec{S}^{(1)}_{n,m}.\vec{S}^{(1)}_{n+1,m} +
\vec{S}^{(2)}_{n,m+{1 \over 2}}.\vec{S}^{(2)}_{n+1,m+{1 \over 2}} \nn
& + & J_{\perp} \sum_{m,n} \vec{S}^{(1)}_{n,m}.
\( \vec{S}^{(2)}_{n,m+{1 \over 2}}+\vec{S}^{(2)}_{n,m-{1 \over 2}}\) \nn
& + & J_{\perp} \sum_{m,n} \vec{S}^{(2)}_{n,m+{1 \over 2}}.
\( \vec{S}^{(1)}_{n+1,m}+\vec{S}^{(1)}_{n-1,m+1}\) \nn
&+& {\cal I}_{\text{\tiny DM}}\, .
\eea
Due to the doubling in the y-direction, the Fourier transform of the
interchain spin-spin couplings is now
\be
J(\vec{k})= 2 J_{\perp}\big( \cos(k_y/2)+ \cos(k_x-k_y/2) \big) \, .
\ee

In order to choose the quantization axes for sets (1) and (2) we
introduce the unit vectors
\bea
&&\vec{u}_1=\vec{D}_1/|\vec{D}_1| \, , \quad \vec{u}_2=\vec{D}_2/|\vec{D}_2| \, , \\
&&\quad \vec{v}={\vec{u}_1\times \vec{u}_2 }/ {| \vec{u}_1\times \vec{u}_2|} \, .
\eea
The spins along the chains of type (1) and (2) will be quantized in
local coordinate systems with axes $(\vec{v},\vec{u}_1\times
\vec{v},\vec{u}_1)$ and $(\vec{v},\vec{u}_2\times \vec{v},\vec{u}_2)$
respectively. We denote the angle between $\vec{u}_1$ and $\vec{u}_2$
by $\zeta$ and define two DM angles by
$\theta_1=-\arctan(|\vec{D}_1|/J_{\parallel})$,
$\theta_2=-\arctan(|\vec{D}_2|/J_{\parallel})$.

Specializing to ${\rm Cs_2CuCl_4}$, we note that the space symmetry
group for this material is $Pnma$. This group contains symmetry
elements which map the spin chains of different types onto each other,
which leads to the following restrictions of the DM vectors
$\vec{D}_1$ and $\vec{D}_2$. They are perpendicular to the chain
direction, perpendicular to each other and of equal length.
This yields that $\theta_1=\theta_2$ in this case; see
\cite{Coldea97b} for a detailed symmetry analysis. 

Next we define the usual step operators for $j=1,2$~:
\be
S^{\pm}_{(j)}=S^x_{(j)} \pm i S^y_{(j)} \, .
\ee
Notice that because of our choice of the quantization axis,
the letters $x,y,z$ refer to different directions for $j=1$ and $j=2$.
In Fourier space, the total Hamiltonian is
\bea
&&\HDM = 
J_{\parallel} \sum_{\vec{k}; \, j=1,2} 
\frac{\cos(k_x-\theta_j)}{\cos{\theta_j}}
S^+_{(j)}(\vec{k})S^-_{(j)}(-\vec{k})  \nn
&&+ J_{\parallel} \sum_{\vec{k}; \, j=1,2} \cos k_x S^z_{(j)}(\vec{k})S^z_{(j)}(-\vec{k}) \nn
&&+\cos\zeta \sum_{\vec{k}} J(\vec{k})\Bigl[
S^y_{(1)}(\vec{k}) S^y_{(2)}(-\vec{k})+S^z_{(1)}(\vec{k}) S^z_{(2)}(-\vec{k})\Bigr] \nn
&&+\sin\zeta \sum_{\vec{k}} J(\vec{k})\Bigl[
S^z_{(1)}(\vec{k}) S^y_{(2)}(-\vec{k})-S^y_{(1)}(\vec{k}) S^z_{(2)}(-\vec{k})\Bigr]\nn
&&+\sum_{\vec{k}} J(\vec{k})  S^x_{(1)}(\vec{k}) S^x_{(2)}(-\vec{k}) \,.
\eea

Next we will write the ``effective'' quadratic spin Hamiltonian
corresponding to the random phase approximation. In order to do so, we
define 
\bea
\Sigma_{j}&=&{1\over 2} \([\chi^{\rm xx}_{j}(k_x+\theta_j)]^{-1}
+[\chi^{\rm xx}_{j}(k_x-\theta_j)]^{-1}\) \, ,\nn
\Delta_{j}&=&{1\over {2i}} \([\chi^{\rm xx}_{j}(k_x+\theta_j)]^{-1}
-[\chi^{\rm xx}_{j}(k_x-\theta_j)]^{-1}\),\nn
\Omega_j&=&[\chi_j^{\rm zz}(k_x)]^{-1}\ .
\eea
Here $\chi^{\rm xx}_j$ and $\chi^{\rm zz}_j$ denote the time-ordered
imaginary-time correlation functions for chain $(j)$ in the presence
of the DM interaction with coupling $\vec{D}_j$. 
They are related to the transverse and longitudinal dynamical
susceptibilities for chain $(j)$ by a Wick rotation and a global
minus sign.

The effective RPA correlation functions between spin operators
$(S^x_1,S^y_1,S^z_1,S^x_2,S^y_2,S^z_2)(\vec{k})$ and spin operators
$(S^x_1,S^y_1,S^z_1,S^x_2,S^y_2,S^z_2)^t(-\vec{k})$ are then given
by the inverse of the bilinear form
\be
\label{matrix}
\left[
\begin{array}{cccccc}
 \Sigma_1  &   \Delta_1    &       0       &      J     &        0      &     0         \\
 -\Delta_1 &   \Sigma_1    &       0       &      0     &  J\cos\zeta & J\sin\zeta  \\
     0     &      0        &     \Omega_1       &      0     & -J\sin\zeta & J\cos\zeta  \\
     J     &      0        &       0       &  \Sigma_2  &    \Delta_2   &     0         \\
     0     & J\cos\zeta  & -J\sin\zeta & -\Delta_2  &    \Sigma_2   &     0         \\
     0     & J\sin\zeta  &  J\cos\zeta &      0     &        0      &    \Omega_2        \\
\end{array}
\right].
\ee
The emergence of a pole in the dynamical susceptibility corresponds to
the vanishing of the determinant of this matrix. This leads to the
following necessary condition 
\bea
0 &=& J^6 -\left[ (1+\cos^2\zeta)\Sigma_1 \Sigma_2+\sin^2\zeta (\Sigma_1 \Omega_2+\Sigma_2 \Omega_1)  \right. \nn
&+& \left. \X1\X2 \cos^2 \zeta  - 2\Delta_1\Delta_2\cos\zeta  \right] J^4 \nn
&+& \left[ (1+\cos^2\zeta)\X1\X2 \Sigma_1 \Sigma_2 \right. \nn
&+& \sin^2\zeta \(\X1 \Sigma_1 (\Sigma_2^2+\Delta_2^2)+\X2 \Sigma_2 (\Sigma_1^2+\Delta_1^2)\)   \nn
&+& \left. \cos^2\zeta (\Sigma_1^2+\Delta_1^2)(\Sigma_2^2+\Delta_2^2)
-2\X1\X2\Delta_1\Delta_2 \cos\zeta \right] J^2 \nn
&-& \X1\X2 (\Sigma_1^2+\Delta_1^2)(\Sigma_2^2+\Delta_2^2) \, .
\eea
In the particular case when $\vec{u}_1=\vec{u}_2$, but $\theta_1$ is
not necessarily equals to $\theta_2$, the condition simplifies to
\bea
0&=&[1-J^2\chi^{\rm zz}_1(k)\chi^{\rm zz}_2(k)]
[1-J^2\chi^{\rm xx}_{1}(k+\theta_1)\chi^{\rm xx}_{2}(k+\theta_2)] \nn
&& \times[1-J^2\chi^{\rm xx}_{1}(k-\theta_1)\chi^{\rm xx}_{2}(k-\theta_2)] \, ,
\eea
where $k$ stands for $k_x$. If in addition to $\vec{u}_1=\vec{u}_2$ we
also have $\theta_1=\theta_2$ then we recover, as we must, the instability
conditions discussed before.

The dynamical susceptibilities can be calculated, by inverting (\ref{matrix}),
then continuing analytically on the frequencies.

\subsection{Dzyaloshinskii-Moriya interaction along the interchain bonds}

Up to this point we have only considered the case when the DM
interaction involves spin along the chains. It was recently suggested
to us by R. Coldea and A. Tennant \cite{radu} that the DM interaction
in ${\rm Cs_2CuCl_4}$ may also involve pairs of spin along the
interchain couplings. 

\begin{figure}[h]
\epsfxsize=7cm
\epsfbox{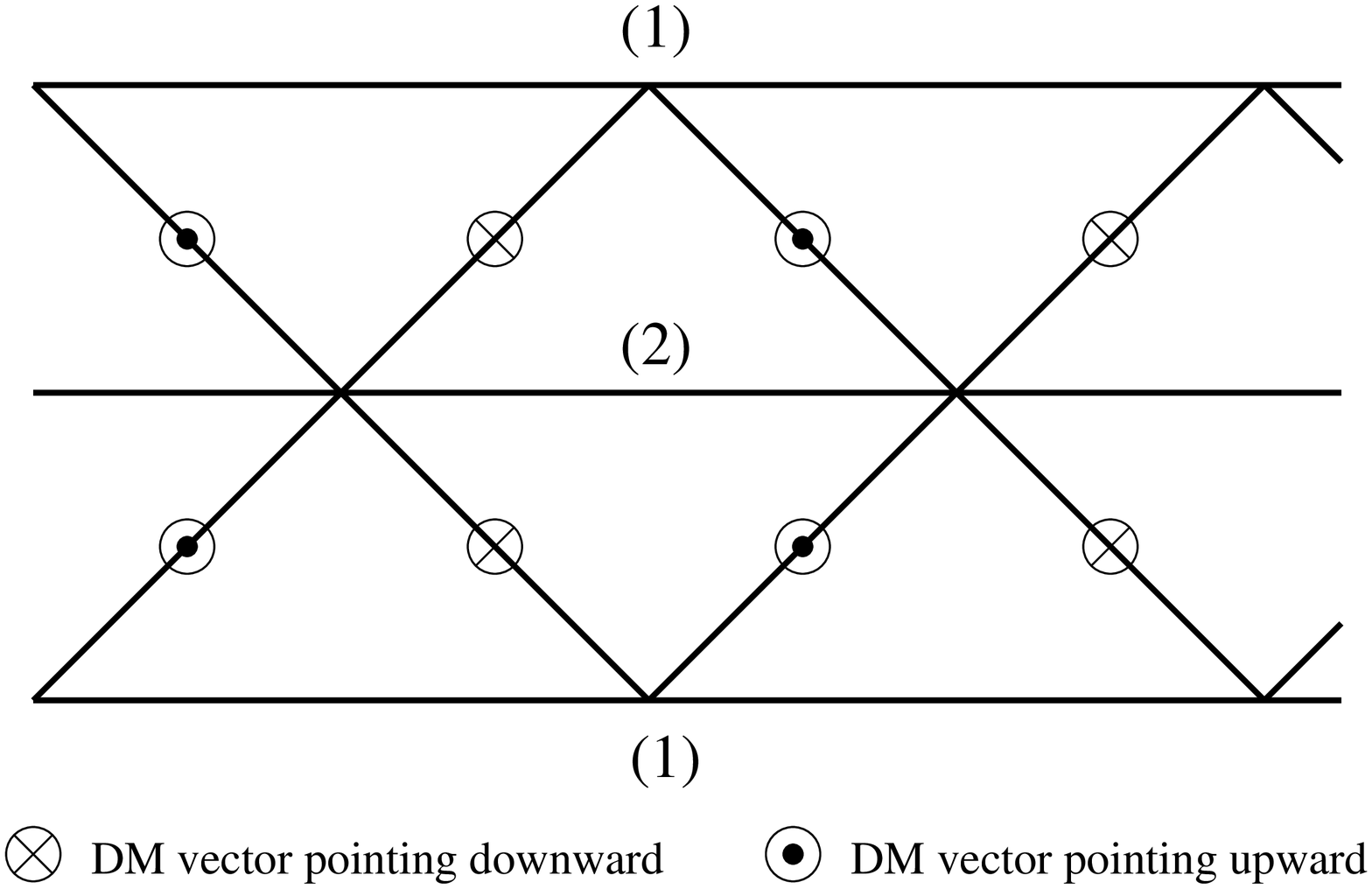}
\caption{Schematic representation of the orientation of the allowed
directions of the DM vector within one layer for the case where the
interaction involves spins along the interchain bonds. } 
\label{fig:dm_2}
\end{figure}

The direction of the DM vector in this case is again constrained by
crystal symmetries \cite{radu}. The DM vector is perpendicular to the
triangular planes and is staggered along the chain direction
as shown in figure (\ref{fig:dm_2}).
For ${\rm Cs_2CuCl_4}$ the DM vector appears to be staggered between
triangular layers as well \cite{radu}. In what follows we will 
first consider the simpler case in which there is no
staggering between layers in order to keep our formulas simple. 

We again need to distinguish between spin chains of type (1) and
(2). The calculations are similar to the ones of subsection \ref{ssec:DM1}.
Since the longitudinal spin operators $S^{z}_{(j)}$ do not
couple to the transverse ones, the transverse and longitudinal
susceptibilities can be calculated separately.
In Fourier space, the Hamiltonian takes the form
\bea
\HDM &=& {\cal H}_0+{\cal H}_{\rm int}\ ,\nn
{\cal H}_0&=&
J_{\parallel} \sum_{\vec{k}; \, j=1,2}\sum_{\alpha} \cos k_x\
S^\alpha_{(j)}(\vec{k})S^\alpha_{(j)}(-\vec{k})  \nn
{\cal H}_{\rm int}&=& \sum_{\vec{k}}\vec{{\cal S}}(\vec{k})^T\
A(\vec{k})\ \vec{{\cal S}}^\dagger(-\vec{k})\nn
&&+ \sum_{\vec{k}} 2\tilde{J}(\vec{k})\ S^z_{(1)}(\vec{k}) S^z_{(2)}(-\vec{k})\nn
&&+ \sum_{\vec{k}}\sum_{j=1,2} J_z\cos k_z\ S^z_{(j)}(\vec{k}) S^z_{(j)}(-\vec{k})\, ,
\eea
where $D$ denotes the DM coupling,
\bea
&&\vec{{\cal S}}(\vec{k})^T=\bigl(S^+_{(1)}(\vec{k}),S^+_{(2)}(\vec{k})\bigr)\ ,\nn
&&\tilde{J}(\vec{k})= J_{\perp}\big( \cos (k_y/2) +
\cos(k_x-k_y/2) \big),\nn 
&&K(\vec{k})= D \big(  \sin(k_y/2) +\sin(k_x-k_y/2)+ \big),
\eea
and $A(\vec{k})$ is the $2\times 2$ matrix
\be
\left[
\begin{array}{cc}
 J_z\cos k_z&\tilde{J}+K     \\
 \tilde{J}+K& J_z\cos k_z     \\
\end{array}
\right].
\ee

It will be convenient to define
\bea
L_{\theta}(\vec{k})&=&{\tilde{J}(\vec{k}) + K(\vec{k})}\nn
&=&\frac{J_{\perp}}{\cos\theta}\big[
\cos(\frac{k_y}{2}+\theta)+ \cos(k_x-\frac{k_y}{2}+\theta) \big],
\eea
where $\theta=-\arctan (D/J_{\perp})$.

We denote by $\chi^{+-}(\vec{k})\equiv\chi^{-+}(\vec{k})$ the
time-ordered imaginary-time transverse correlation function between
spins $S^+_{(j)}(\vec{k})$ and spin $S^-_{(j)}(-\vec{k})$ in the
absence of the interchain coupling and DM interaction. This is of
course simply the correlation function of a single one dimensional
chain. 

The RPA expressions for the transverse correlation functions between
spin operators ${\cal S}_\alpha(\vec{k})$ and
${\cal S}^\dagger_\beta(-\vec{k})$
($\alpha,\beta=1,2$) are given by the matrix
elements of the inverse of the following $2\times 2$ matrix

\be
\left[
\begin{array}{cc}
 [\chi^{+-}]^{-1}+J_z \cos k_z&L_{\theta}\\
 L_{\theta}&{[\chi^{+-}]}^{-1}+J_z \cos k_z \\
\end{array}
\right].
\label{shiftDM2}
\ee

The time ordered imaginary time
two point correlation function of spins is obtained by adding the
contributions from the various sublattice correlators, i.e. by taking
e.g.
\be
\frac{1}{2}\sum_{j,l}\langle S^+_{(j)}(\vec{k})S^-_{(l)}(-\vec{k})\rangle .
\ee

After analytic continuation we obtain the following RPA formula for
the transverse dynamical susceptibilities 
\bea
\label{chi3dDM_2}
\chi^{+-}_{3d}(\omega,\vec{k})&=&
\frac{\chi^{+-}(\omega,k_x)}{1-[L_{\theta}(\vec{k})+J_z\cos k_z]\;\chi^{+-}(\omega,k_x) } \, , \nn
\chi^{-+}_{3d}(\omega,\vec{k})&=& \chi^{+-}_{3d}(\omega,-\vec{k})\, .
\eea
Here $\chi^{+-}(\omega,k_x)$ is the dynamical susceptibility of a
single one dimensional chain.
We note that we could have arrived at this result by first removing
the DM interaction along the interchain bonds by means of
a unitary transformation in a way analogous to \r{trafo}. This induces
an effective DM interaction along the chains and the resulting
Hamiltonian can again be analyzed within RPA. After undoing the
unitary transformation one recovers \r{chi3dDM_2}.

The structure of the dynamical susceptibilities (\ref{chi3dDM_2}) is quite
similar to the one we obtained in the case where the DM interaction is
along the chain direction \r{chi3dDM}. There again is an effective
global shift of the momentum by a ``DM angle'' $\theta$.
However, now the angle depends on the ratio of the DM coupling $D$ to
the interchain coupling $J_{\perp}$ rather than the strong exchange
$J_{\parallel}$, which makes it much larger. This leads to an
enhancement of the effect of the DM interaction.

The condition for the emergence of a transverse instability towards
the formation of a coherent mode reads
\be
1=\max_{\vec{k}}\left\lbrace [L_{\theta}(\vec{k})+J_z\cos k_z]\ 
\chi^{+-}(\omega,k_x)\right\rbrace .
\ee
One finds that the instability again occurs along the chain direction,
i.e. $k_y=k_x$. 

\subsection{Taking into account the staggered layers}

It was suggested in \cite{radu} that in ${\rm Cs_2CuCl_4}$
the direction of the DM vector alternates between neighbouring layers.
This feature can be accomodated within our calculation as follows.

Let us first suppose that the interlayer coupling constant $J^z$ is
negligible. Then the global transverse dynamical susceptibilities are
given by the sums over the contributions from the two types of
layers. In this case the transverse susceptibility reads
\bea
\label{chi3dDM_3}
&& \chi^{+-}_{3d}(\omega,\vec{k}) = \nn
&& \frac{\chi^{+-}(\omega,k_x)\(1-J(\vec{k})\,\chi^{+-}(\omega,k_x)\)}
{\(1-L_{\theta}(\vec{k})\;\chi^{+-}(\omega,k_x)\)
\(1-L_{-\theta}(\vec{k})\;\chi^{+-}(\omega,k_x)\) } \, , \nn
&&\chi^{-+}_{3d}(\omega,\vec{k})= \chi^{+-}_{3d}(\omega,\vec{k})\, .
\eea
As expected the transverse dynamical susceptibility has lost its
chiral nature, i.e. we now have $\chi^{-+}_{3d}= \chi^{+-}_{3d}$.
This is in marked contrast to the case where the DM interaction
involves spins along the chains, for which we always had
$\chi^{-+}_{3d} \neq \chi^{+-}_{3d}$.

If $J_z$ is not zero the calculations are slightly more complicated. 
One has now to distinguish spins in neighbouring layers and the
elementary cell is doubled in the z-direction. We will denote the spin
operators corresponding to the four sites per unit cell by
$S^\alpha_{(j,k)}$ ($j,k=1,2$). Let us also introduce the function
$I(\vec{k})=J^z\cos(k_z/2)$. 
Then the effective RPA transverse correlation
functions between spin operators
$(S^+_{(1,1)},S^+_{(2,1)},S^+_{(1,2)},S^+_{(2,2)})(\vec{k})$ 
and
$(S^-_{(1,1)},S^-_{(2,1)},S^-_{(1,2)},S^-_{(2,2)})^T(-\vec{k})$
are given by the inverse of the following bilinear form
\be
\left[
\begin{array}{cccc}
 [\chi^{+-}]^{-1} & L_{\theta}  &  I    &            0\\
 L_{\theta}     &   [\chi^{+-}]^{-1}   &          0            &        I      \\
       I        &            0             & [\chi^{+-}]^{-1}  &   L_{-\theta} \\
           0             &        I        & L_{-\theta}  &   [\chi^{+-}]^{-1} 
\end{array}
\right].
\ee

The time ordered imaginary time two point
function of spin operators is again obtained by summing over the
sublattice contributions. After analytic continuation, we obtain the
following RPA expression for the transverse dynamical susceptibility
\bea
\label{chi3dDM_4}
&& \chi^{+-}_{3d}(\omega,\vec{k}) = \nn
&& \frac{\chi^{+-}(\omega,k_x)\(1+N_1(\vec{k})\ 
\chi^{+-}(\omega,k_x)\)}
{\(1-2\tilde{J}(\vec{k}) \chi^{+-}(\omega,k_x)+N_2(\vec{k})\
\left[\chi^{+-}(\omega,k_x)\right]^2\)} \ ,
\eea
where
\bea
N_1(\vec{k})&=&I(\vec{k})-\tilde{J}(\vec{k})\ ,\nn
N_2(\vec{k})&=&\tilde{J}^2(\vec{k})-K^2(\vec{k})-I^2(\vec{k}).
\eea

We again have that
\be
\chi^{-+}_{\rm 3d}(\omega,\vec{k})= \chi^{+-}_{\rm 3d}(\omega,\vec{k}) \, .
\ee
From (\ref{chi3dDM_4}) we obtain a modified set of instability
conditions
\be
\(J(\vec{k})\pm \sqrt{K^2(\vec{k})+I^2(\vec{k})}\)\chi^{+-}(0,k_x)=1 \, .
\ee
Extremizing with respect to $k_z$ and $k_y$ yields the conditions
$k_y=k_x$ and $k_z=0$.

If we specify the exchange couplings according to the values suggested
for Cs$_2$CuCl$_4$ ($J_z \simeq 0.05 J_{\parallel}$,
$J_\perp \simeq 0.33 J_{\parallel}$, $ D \simeq 0.05
J_{\parallel}$) we obtain a critical temperature of $T_c=0.727$ K 
and an ordering wave number of $k_0=0.154$. These are close to the
experimental values $T_c=0.62$ K and $k_0=0.186$.

\begin{center}
\begin{tabular}{|c||c|c|c|}
\hline
$\theta$
& ${J_\perp}/{J_\parallel}$ =0.1
& ${J_\perp}/{J_\parallel}$ =0.2
& ${J_\perp}/{J_\parallel}$ =0.3\\
\hline\hline
0.05
& $t_c=0.070$
& $t_c=0.078$
& $t_c=0.092$\\
&  $k_0=0.021$
&  $k_0=0.046$
&  $k_0=0.082$\\
\hline
0.10
& $t_c=0.074$
& $t_c=0.091$
& $t_c=0.120$\\
&  $k_0=0.022$
&  $k_0=0.053$
&  $k_0=0.103$\\
\hline
0.15
& $t_c=0.079$
& $t_c=0.109$
& $t_c=0.153$\\
&  $k_0=0.023$
&  $k_0=0.062$
&  $k_0=0.128$\\
\hline
0.20
& $t_c=0.087$
& $t_c=0.130$
& $t_c=0.189$\\
&  $k_0=0.025$
&  $k_0=0.073$
&  $k_0=0.153$\\
\hline
\end{tabular}
\end{center}
\underline{Table 4:} {\small Transition temperatures
$t_c=T_c/J_\parallel$ and ordering wave numbers $k_0$ for various values
of the frustrated ($J_\perp$) interchain coupling and interchain DM angle
$\theta=\arctan(D/J_\perp)$. The interplane coupling $J_z$ is taken to be $0.05 J_{\parallel}$.
}

%\subsection{Effect of a magnetic field}
%Let us briefly discuss the effects of applying a uniform magnetic
%field $H$ in the presence of a staggered DM interaction. As before we
%choose the DM vector $\vec{D}$ to point along the z-direction in spin space.
%
%\subsubsection{$\vec{H}$ parallel to $\vec{D}$}
%
%\subsubsection{$\vec{H}$ at an angle to $\vec{D}$}

\section{Summary and Conclusions}
\label{sec:summ}
In this work we have studied the dynamical response of frustrated
quasi-1D spin-1/2 Heisenberg magnets in the disordered phase. The
starting point of our approach are exact results for the
finite-temperature dynamical susceptibility of an ensemble of
uncoupled chains. Taking the couplings between chains into account
within the framework of a random phase approximation, we obtained an
analytic expression for the dynamical structure factor of the quasi-1D
system we are interested in. In the disordered phase the main effects
of the frustration are to generate an {\sl asymmetry} of the line
shape and a shift of the apparent dispersion to an incommensurate wave
vector. 

By analyzing the instability of the disordered phase with respect to
the formation of collective modes we studied the transition
to the low-temperature ordered phase. In particular we determined
the ordering temperature and ordering wave vector within the framework of
our approach. We found that there is a very weak instability
towards an incommensurately ordered state.

We then considered the effects of an applied magnetic field. We found
that for isotropic Heisenberg magnets application of a magnetic field
leads first to an increase in the transition temperature $T_c(H)$ and for very
large field to an eventual decrease. 
In the presence of an exchange anisotropy we found two
distinct behaviours: if the field is applied along the direction of
the anisotropy the situation is very similar to the isotropic
case. Applying the field at an angle to the anisotropy generates a gap
in the individual chains, which leads to a decrease of $T_c(H)$ with
$H$ in the quasi-1D system. This type of magnetic phase
diagram is qualitatively similar to what has been observed in ${\rm
Cs_2CuCl_4}$. 

In the second part of this work we took into account the effects of
various types of Dzyaloshinskii-Moriya interactions, which break spin
rotational invariance. We first considered DM interactions involving
spins along the chain directions and then DM interactions involving
spins along interchain bonds. 
In the disordered phase the main effect of DM interactions is to
generate stronger incommensurations. DM interactions also leads to a
significant enhancement in the transition temperature and ordering
wave number. 

In general our results are qualitatively similar to what is observed
experimentally in ${\rm Cs_2CuCl_4}$. In presence of a DM interaction
of the kind proposed in \cite{radu} for ${\rm Cs_2CuCl_4}$ we obtain a
transition temperature and ordering wave number, which are close to
the experimental values. However, given that the interchain coupling
is not small it is unclear how reliable the RPA approach is. In light
of this fact it would be very interesting to determine the leading
corrections to RPA, which can be done for example along the lines of
\cite{irkhin}. It also would be interesting to extend the coupled
chains approach to the ordered phase. This is not straightforward due
to the presence of ``twist''-operators \cite{NGE}.

{\bf Acknowledgements:} 
We thank Radu Coldea and Alan Tennant for many stimulating
discussions and explaining their experimental results to us.
This work was supported by 
the EPSRC under grants AF/100201 (FHLE) and GR/N19359 (MB and FHLE). 
% and to E. Barrios for helpful correspondence.

\end{document}